\documentclass[aps,prx,twocolumn,floatfix,10pt,superscriptaddress]{revtex4-2}
\usepackage[utf8]{inputenc}
\usepackage{amsmath,amsfonts,hyperref,dsfont,subcaption,relsize,placeins}
\usepackage[normalem]{ulem}
\usepackage{ragged2e}
\usepackage[version=4]{mhchem}
\usepackage{xcolor,xfrac}
\usepackage{tikz}
\usepackage{floatrow}
\usepackage{hyperref}
\usepackage{soul}
\usetikzlibrary{quantikz}
\hypersetup{colorlinks=true,
						linkcolor=blue,      
						citecolor=blue,            
						filecolor=blue,             
						urlcolor=blue              
}
\DeclareCaptionJustification{justified}{\justifying}

\begin{document}
\title{Nonlinear dynamics as a ground-state solution on quantum computers}
\author{Albert J. Pool}
\author{Alejandro D. Somoza}
\affiliation{Institute of Engineering Thermodynamics, German Aerospace Center (DLR), Wilhelm-Runge-Str.\ 10, 89081 Ulm, Germany}
\affiliation{Helmholtz Institute Ulm, Helmholtzstr.\ 11, 89081 Ulm, Germany}
\author{Conor \surname{Mc Keever}}
\affiliation{Quantinuum, Partnership House, Carlisle Place, London SW1P 1BX, United Kingdom}
\author{Michael Lubasch}
\affiliation{Quantinuum, Partnership House, Carlisle Place, London SW1P 1BX, United Kingdom}
\author{Birger Horstmann}
\email{michael.lubasch@quantinuum.com, birger.horstmann@dlr.de}
\affiliation{Institute of Engineering Thermodynamics, German Aerospace Center (DLR), Wilhelm-Runge-Str.\ 10, 89081 Ulm, Germany}
\affiliation{Helmholtz Institute Ulm, Helmholtzstr.\ 11, 89081 Ulm, Germany}
\affiliation{Department of Physics, Ulm University, Albert-Einstein-Allee 11, 89081 Ulm, Germany}

\begin{abstract}
For the solution of time-dependent nonlinear differential equations, we present variational quantum algorithms (VQAs) that encode both space and time in qubit registers.
The spacetime encoding enables us to obtain the entire time evolution from a single ground-state computation.
We describe a general procedure to construct efficient quantum circuits for the cost function evaluation required by VQAs.
To mitigate the barren plateau problem during the optimization, we propose an adaptive multigrid strategy.
The approach is illustrated for the nonlinear Burgers equation.
We classically optimize quantum circuits to represent the desired ground-state solutions, run them on IBM Q System One and Quantinuum System Model H1, and demonstrate that current quantum computers are capable of accurately reproducing the exact results.
\end{abstract}

\maketitle

\section{Introduction}
Methods to solve partial differential equations (PDEs), particularly those which are nonlinear, have long been of central importance in fields such as aerospace engineering~\cite{mani2023perspective} and energy science~\cite{porte2020wind}.
Prominent applications include solving the Navier--Stokes equation in computational fluid dynamics and the numerical integration of continuum models in battery research~\cite{schammer2021theory,bolay2022microstructure}.
Despite their efficacy, conventional numerical methods encounter substantial constraints when tackling large-scale three-dimensional models~\cite{tadmor2012review}.

Numerous quantum algorithms have been proposed to integrate PDEs with an exponential advantage over their classical counterparts in theory~\cite{GiviEtAl20}.
This includes quantum linear systems algorithms (QLSAs) for linear~\cite{cao_quantum_2013,berry2014high,montanaro_quantum_2016,BeEtAl17,CoJoOs19,childs2021high,saha2023advancing,cui_quantum_2023,bharadwaj_hybrid_2023,ingelmann_two_2023} and nonlinear~\cite{lloyd2020quantum,liu2021efficient,jin2022quantum,lapworth2022hybrid,JiLiJu23} PDEs, as well as quantum algorithms based on Hamiltonian simulation~\cite{wiesner1996simulations,zalka1998simulating,CoJoOs19,arrazola2019quantum,lloyd2020quantum,JiLiYu22a,JiLiYu22b,JiLiJu23,wright2024noisy}.
Furthermore, techniques based on quantum amplitude and phase estimation have been put forth to integrate PDEs~\cite{XuEtAl18,XuEtAl19,gaitan2020finding,gaitan2021finding,oz2022solving} (which might be realizable using approximations with low quantum hardware requirements~\cite{SuEtAl20,rao2020quantum,PlEtAl22}).
In practice, however, these algorithms do not yet give a quantum advantage.
Various challenges, including the input-output problem, hinder an actual quantum advantage in some cases~\cite{aaronson2015read,linden2022quantum}.
On current noisy intermediate-scale quantum (NISQ) devices~\cite{preskill2018quantum}, these algorithms are affected by strong limitations on the depth of the quantum circuits that can be coherently executed. 

On NISQ devices, variational quantum algorithms (VQAs) have become a popular approach to solving PDEs~\cite{cerezo2021variational}.
One famous example is the Variational Quantum Linear Solver, which is a generic solver for linear systems of equations and can be applied to linear PDEs~\cite{liu2022application,demirdjian2022variational,liu_variational_2024}.
A standard benchmark is the Poisson equation~\cite{liu2021variational,PhysRevA.104.052409,joo2021quantum,guseynov2023depth,li2023variational}.
For nonlinear equations,
different algorithmic primitives have been proposed~\cite{mezzacapo_quantum_2015,lubasch2020variational,budinski2021quantum,kyriienko2021solving,paine2023quantum}.

One way of solving time-dependent PDEs using VQAs proceeds one time step at a time, analogous to established classical numerical solvers~\cite{li_numerical_2017}.
In this time-stepping approach, however, a large number of time steps can lead to an accumulation of errors. Also note that time stepping leads to a computational complexity that is at least linear in the number of time steps.

\begin{figure*}
    \centering
    \includegraphics[scale=0.76]{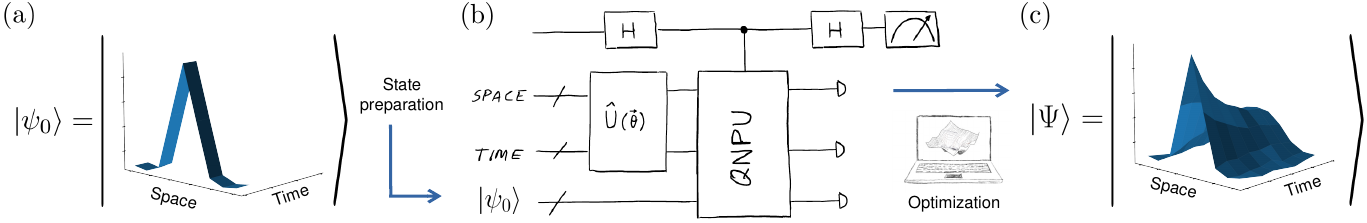}
    \caption{Schematic summarizing the proposed approach.
    (a) The quantum state $\ket{\psi_0}$ encodes the initial condition at $t=0$.
    (b) The nonlinear Hamiltonian is evaluated using QNPUs. The solution to the PDE at all times corresponds to the ground state of the nonlinear Hamiltonian, which can be calculated via the variational optimization of an ansatz $U(\vec{\theta})$ with $n_x+n_t$ qubits and variational parameters $\vec{\theta}$. (c) 3D plot of the solution corresponding to the result from the Quantinuum H1-1 computer. See Fig.~\ref{fig2:burgers_quantinuum} for more details.
    }
    \label{fig1:schematic_method}
\end{figure*}

In this article, our goal is to develop alternative VQAs that do not rely on time stepping. To that end, we use the formalism of Feynman~\cite{feynman_simulating_1982,feynman_quantum_1985} and Kitaev~\cite{kitaev_classical_2002}, in which time is encoded in a clock qubit register and the solution to a time-dependent problem at all points in time is contained in the ground state of a Hamiltonian.
A proof-of-principle demonstration of this \textit{spacetime} approach was introduced by McClean \textit{et al}.\ for quantum chemical problems in 2013~\cite{jarrod_mcclean_feynmans_2013}, which was extended to open quantum systems by Tempel \textit{et al.}\ in 2014~\cite{tempel2014kitaev}. A recent implementation of this method has been put forth by Barison \textit{et al.}~\cite{barison2022variational} for the dynamics of quantum systems, providing numerical evidence for a favorable scaling.

We present methods that extend the Feynman--Kitaev formalism to incorporate nonlinear dynamics~\cite{pool2022solving}. 
We show how to evaluate the Feynman--Kitaev Hamiltonian using quantum nonlinear processing units (QNPUs)~\cite{lubasch2020variational,jaksch2023variational,sarma_quantum_2024}.
In order to mitigate barren plateaus, which are a serious problem for variational algorithms~\cite{mcclean2018barren},
and scale the variational algorithm to a larger number of qubits, we propose a multigrid optimization strategy with a customized ansatz structure~\cite{lubasch2018multigrid}.
In this approach, a converged ansatz is passed from a coarser to a finer grid, which allows us to speed up the optimization by limiting the number of variational parameters.
We apply the approach to the Burgers equation as a paradigmatic nonlinear equation. We show that the number of circuits required to evaluate the Feynman--Kitaev Hamiltonian only depends on the number of terms in the equation and the order of approximation in the time step, not on the number of time steps or spatial points, and that these circuits have a depth that scales at most linearly with the number of qubits.
A schematic of the approach is depicted in Fig.~\ref{fig1:schematic_method}.

This article has the following structure.
In Sec.~\ref{sec:method}, we first present the modified Feynman--Kitaev cost function and its implementation in terms of QNPUs. We then show how to evaluate this cost function on NISQ devices and how to optimize it using VQAs.
In Sec.~\ref{sec:results}, we present results from the IBMQ and Quantinuum devices. We then show the cost achieved for different variational ansatz structures and discuss the scalability of the algorithm. We provide our conclusions in Sec.~\ref{sec:conclusion} and technical details in the Appendices.

\section{Methods}
\label{sec:method}
Here we describe how we use the Feynman--Kitaev formalism to solve nonlinear PDEs on NISQ devices.
In Sec.~\ref{sec:method_A}, we derive the Feynman--Kitaev Hamiltonian for nonlinear systems. In Sec.~\ref{sec:method_B}, we show how to construct QNPU circuits to evaluate it. In Sec.~\ref{sec:method_C}, we describe the variational procedure which we use to find the ground state of the Hamiltonian. Finally, we discuss in Sec.~\ref{sec:method_D} what is necessary to implement this Hamiltonian on a NISQ device.

\subsection{Feynman--Kitaev Hamiltonian for nonlinear systems}
\label{sec:method_A}
We want to apply the Feynman--Kitaev formalism to time-dependent differential equations of the form
\begin{equation}\label{eq:diffeq}
 \frac{d}{dt}f(x,t) = \mathcal{L}[f(x,t)]f(x,t),
\end{equation}
where $\mathcal{L}$ is a (linear or nonlinear) differential operator acting on the function $f(x,t)$. To implement nonlinear equations, the operator $\mathcal{L}$ may have a functional dependence on $f(x,t)$.
Such an equation can be solved classically by defining a propagator~\cite{li_numerical_2017}
\begin{equation}\label{eq:main-propagator}
 \hat{T}(d t) = e^{dt\, \mathcal{L}}.
\end{equation}
The exponential of the operator can then be approximated using the Taylor series
\begin{equation}\label{eq:T}
 \hat{T}(dt) = \mathds{1} + dt \, \mathcal{L} + (dt \, \mathcal{L})^2/2  + \mathcal{O}(dt^3),
\end{equation}
where the time step is assumed to be small, i.e., $dt \ll 1$. Although this is formulated for a time-independent problem, the generalization to time-dependent problems using a time-ordered propagator is straightforward \cite{jarrod_mcclean_feynmans_2013}.
In contrast to quantum dynamics \cite{barison2022variational}, the propagator $\hat{T}$ is not unitary for classical systems in general.

\begin{figure*}[htb!]
\centering
\begin{subfigure}{.25\textwidth}
    \centering
    \caption{}\vspace{-2em}
    \begin{quantikz}[thin lines,row sep=0.15cm,column sep=0.4cm,font={\small}]
        \lstick{$|0\rangle$} & \gate[2,style={fill=red!20}]{U} \gategroup[6,steps=2,style={dashed, rounded corners, red}] \qw & \qw & \qw \ldots \\
        \lstick{$|0\rangle$} & \qw & \gate[2,style={fill=red!20}]{U} & \qw \ldots \\
        \lstick{$|0\rangle$} & \gate[2,style={fill=red!20}]{U} & \qw & \qw \ldots \\
        \lstick{$|0\rangle$} & \qw & \gate[2,style={fill=red!20}]{U} & \qw \ldots \\
        \lstick{$|0\rangle$} & \gate[2,style={fill=red!20}]{U} & \qw & \qw \ldots \\
        \lstick{$|0\rangle$} & \qw & \qw & \qw \ldots
    \end{quantikz}
    \label{fig1:bw-circ}
\end{subfigure}~
\begin{subfigure}{0.33\textwidth}
    \centering
    \caption{}
    \begin{quantikz}[thin lines,row sep=0.1cm, column sep=0.4cm,font=\small]
    \lstick{\ket{0}} & \gate[3,style={fill=blue!20}]{U} & \qw         & \qw         & \qw         & \qw \\
    \lstick{\ket{0}} & \qw         & \gate[3,style={fill=blue!20}]{U} & \qw         & \qw         & \qw \\
    \lstick{\ket{0}} & \qw         & \qw         & \gate[3,style={fill=blue!20}]{U} & \qw         & \qw \\
    \lstick{\ket{0}} & \qw         & \qw         & \qw         & \gate[3,style={fill=blue!20}]{U} & \qw \\ 
    \lstick{\ket{0}} & \qw         & \qw         & \qw         & \qw         & \qw \\
    \lstick{\ket{0}} & \qw         & \qw         & \qw         & \qw         & \qw
    \end{quantikz}
    \label{fig1:mps-chi-4}    
\end{subfigure}~
\begin{subfigure}{.27\textwidth}
    \centering
    \caption{}
    \begin{quantikz}[thin lines,row sep=0.15cm,column sep=0.4cm,font=\small]
        \lstick[3]{$t$\scalebox{1.5}{$\Big\uparrow$}} & \lstick{$t_2$} & \gate[2,style={fill=red!20}]{U} & \qw & \qw \\
         & \lstick{$t_1$} & \qw & \gate[2,style={fill=red!20}]{U} & \qw \\
         & \lstick{$t_0$} & \gate[2,style={fill=red!20}]{U} & \qw & \qw \\
        & \lstick[3]{$x$\scalebox{1.5}{${\color{blue}\Big\uparrow}{\color{red}\Big\downarrow}$}} & \qw & \gate[2,style={fill=red!20}]{U} & \qw \\
        & & \gate[2,style={fill=red!20}]{U} & \qw & \qw \\
        & & \qw & \qw & \qw
    \end{quantikz}
    \label{fig1:sequential}
\end{subfigure}

    \vspace{1em}
    \begin{subfigure}{0.46\textwidth}
        \centering
        \caption{}\vspace{-2em}
            \begin{quantikz}[thin lines,font=\small,row sep=0.1cm,column sep=0.3cm]
            \lstick{\ket{0}} & \qw & \gate{H} & \qw & \ctrl{2} & \gate{H} & \meter{} \\
            \lstick{\ket{0}} & \qw & \qw & \targ{} & \qw & \qw & \meter{} \\
            \lstick{\ket{0}} & \qwbundle{n_t} & \gate[2,style={fill=yellow!20}]{\hat{U}(\vec{\theta})} & \octrl{-1} & \gate[2,style={fill=green!20}]{\text{QNPU}(\vec{\theta})} & \qw & \qw \\
            \lstick{\ket{0}} & \qwbundle{n_x} & & \qw & & \qw & \qw
        \end{quantikz}
        \label{fig:qnpu-circ-c1}
    \end{subfigure}
    ~
    \begin{subfigure}{0.46\textwidth}
    \centering
    \caption{}\vspace{-2em}
    \begin{quantikz}[thin lines,font=\small,row sep=0.1cm,column sep=0.3cm]
        \lstick{\ket{0}} & \qw & \gate{H} & \ctrl{2} & \ctrl{1} & \gate{H} & \meter{}\\
        \lstick{\ket{0}} & \qw & \qw & \qw & \gate[2]{\hat{A}} & \qw & \qw \\
        \lstick{\ket{0}} & \qwbundle{n_t} & \gate[2,style={fill=yellow!20}]{\hat{U}(\vec{\theta})} & \gate[2,style={fill=green!20}]{\textrm{QNPU}(\vec{\theta})} & & \qw & \qw \\
        \lstick{\ket{0}} & \qwbundle{n_x} & \qw & \qw & \qw & \qw & \qw
    \end{quantikz}
    \label{fig:qnpu-circ-c2}
\end{subfigure}
\caption{Parametrizable quantum circuits employed throughout this work.
{(a)} One layer of a brickwall ansatz, which can be repeated to increase the depth of the circuit. Each unitary $U$ (red) represents a generic two-qubit gate.
{(b)} Generic quantum MPS with bond dimension $\chi=4$ (Sec.~\ref{sec:method_C})
where each unitary $U$ (blue) represents a generic three-qubit gate. 
{(c)} Sequential (blue) and reversed space (red) ordering of space and time qubits in the register. The arrows point from the most to the least significant qubit for time and space.
(d-e) Example QNPU circuits to calculate the cost function, using the adder circuit $\hat A$~\cite{lubasch2020variational} for the time shift operator, where $\hat{U}(\vec{\theta})$ is the ansatz circuit and $H$ the Hadamard gate. Of these, (d) is used for the $\hat{C}_1$ term and (e) for the $\hat{C}_2$ term. In these two figures, the notation $\text{QNPU}(\vec{\theta})$ is used as a short-hand for the QNPU itself and any duplicates of the ansatz, which depends on $\vec{\theta}$, encoded on ancilla qubits. See Appendix \ref{sec:suppl-evaluation} for details about the implementation of these circuits.}\label{fig2}
\end{figure*}
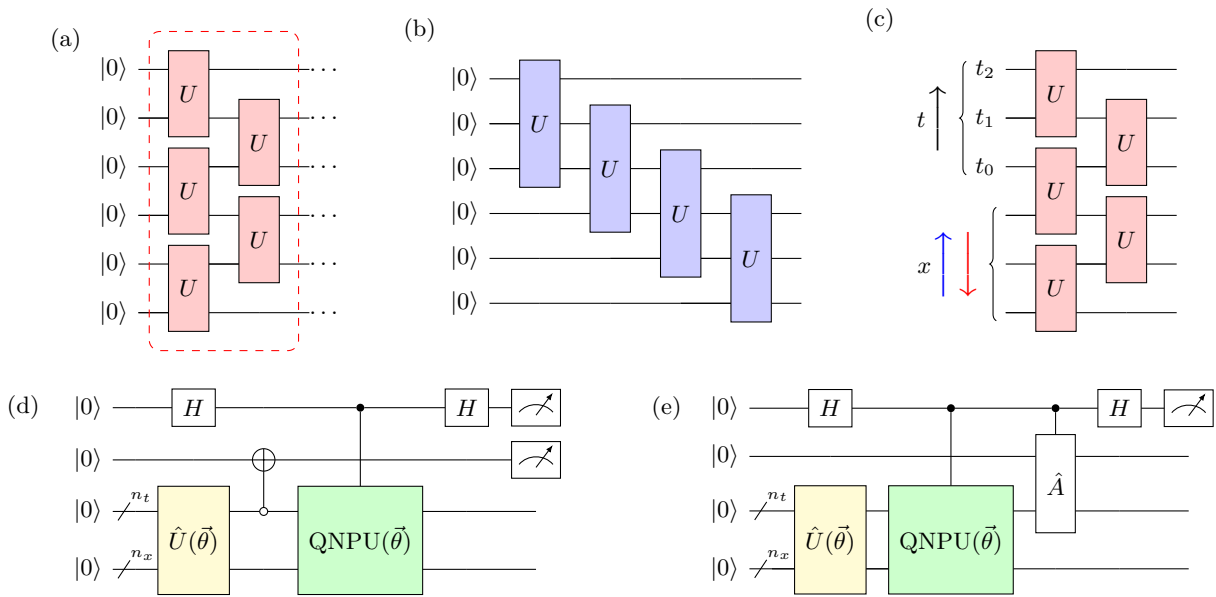

To encode the function $f(x,t)$ in a quantum state $|\Psi\rangle$, we define a spatial register of $n_x$ qubits and a time register of $n_t$ qubits. We then discretize the function $f(x,t)$ on a grid of $2^{n_x}=N_x+1$ points in space and $2^{n_t}=N_t+1$ in time, and use the amplitude encoding~\cite{zalka1998simulating,barison2022variational} given by
\begin{equation}\label{eq:amplitude}
    |\Psi\rangle = \sum_{i=0}^{N_x}\sum_{j=0}^{N_t}\psi_{i,j}|\text{binary}(i)\rangle\otimes|\text{binary}(j)\rangle,
\end{equation}
where $\text{binary}(i)$ is the binary encoding of the number $i$ as a bitstring on the space or time qubits.
This quantum state is normalized as $\sum_{i,j}|\psi_{i,j}|^2=1$.
We now define the Feynman--Kitaev Hamiltonian,
\begin{equation}\label{eq:H}
    \hat{H}=c_0\hat{C}_0 + \hat{X}^\dagger\hat{X},
\end{equation}
where the operator $\hat{C}_0$, which enforces the initial condition $\ket{\psi_0}=f(x,t_0)/\mathcal{N}$, is given by
\begin{equation}\label{eq:C0}
    \hat{C}_0=\left[\hat{I}-|\psi_0\rangle\langle\psi_0|\right]\otimes|0\rangle\langle 0|.
\end{equation}
The constant $c_0$ can be tuned to speed up optimization; the ground state of the Hamiltonian, for which $\langle\hat{C}_0\rangle=0$, does not depend on it. We use $c_0=2$ in all cases presented in this work.
The normalization constant $\mathcal{N}=\|f(x,t_0)\|_{2}$ is used to rescale the results such that the norm of the first time point matches $\mathcal{N}$ and the initial condition is obeyed.
The operator $\hat{X}$, which describes the time evolution, is given by
\begin{equation}\label{eq:X}
    \hat{X} = {\sum_{i = 0}^{N_{t}-1}} \hat{T}(-dt)\otimes |i\rangle \langle i+1| - \hat{I}\otimes |i\rangle \langle i|.
\end{equation}
This $\hat{X}\approx dt (\partial_{t} - \mathcal{L})$ describes an implicit (backward Euler) time integration scheme, known to be more stable for stiff PDEs than explicit schemes \cite{li_numerical_2017}. 
The product $\hat{X}^\dagger \hat{X}$ in Eq.~\eqref{eq:H} can be further decomposed into $\hat{C}_1-\hat{C}_2$ with
\begin{align}
    \hat{C}_1 &= \sum_{i = 0}^{N_{t}-1} \hat{I}\otimes |i\rangle \langle i| + \hat{T}^{\dag}(-dt)\hat{T}(-dt)\otimes |i+1\rangle \langle i+1|, \label{eq:main-c1} \\
    \hat{C}_2 &= \sum_{i = 0}^{N_{t}-1} \hat{T}(-dt) \otimes |i\rangle \langle i+1| + \mathrm{h.c.} \label{eq:main-c2}
\end{align}
We now integrate the PDE of Eq.~\eqref{eq:diffeq} for all time steps by finding the zero-energy ground state of $\hat{H}$. This ground state is given by the \textit{history state}
\begin{equation}\label{eq:history}
\ket{\Psi} = \frac{1}{\sqrt{N_t+1}} \sum_{i=0}^{N_t} [\hat{T}(dt)]^i \ket{\psi_0} \otimes \ket{i}.
\end{equation}
We show numerically in Appendix~\ref{sec:derivation} that the history state is indeed the ground state and that it is non-degenerate. Further details on the derivation of the cost function can also be found there.

\subsection{Evaluation of the Hamiltonian}
\label{sec:method_B}
We evaluate the Feynman--Kitaev Hamiltonian $\langle\hat{H}\rangle$ described in the previous section using the quantum nonlinear processing unit (QNPU) formalism introduced by Lubasch \emph{et al}.~\cite{lubasch2020variational}. 
A QNPU implements nonlinear terms by pointwise multiplication. Lubasch \emph{et al.} present an efficient implementation of spatial derivatives and periodic boundary conditions using a quantum \textit{adder} circuit~\cite{lubasch2020variational}. Nonlinear terms such as $f(x,t)\frac{\partial f(x,t)}{\partial x}$ in the Burgers equation~\eqref{eq:burgers} are represented by combining these two concepts with a duplicate of the ansatz.

We go beyond the approach of Lubasch \emph{et al.}~\cite{lubasch2020variational} by including the time register.
Because our problem is not periodic in time, we add a time qubit to the register before applying the adder circuit $\hat{A}$ for the time derivative, so the time does not wrap around from $|N_t\rangle$ to $|0\rangle$ (see Fig.~\ref{fig:qnpu-circ-c2}).
Using QNPUs, we can estimate the expectation value $\langle\hat{H}\rangle$ in a number of circuits that does not depend on the number of discretization points, with a depth linear in the number of space and time qubits. In this way, we ensure the scalability of the QNPU implementation. 
The number of circuits to realize $\hat{T}$ and its product $\hat{T}^\dag \hat{T}$ as they appear in Eqs.~\eqref{eq:main-c1} and \eqref{eq:main-c2} depends on the number of terms $\mathcal{L}=\mathcal{A}+\mathcal{B}+\ldots$ in the PDE and the powers of $\mathcal{L}$ in the Taylor expansion of the time evolution operator $\hat{T}$ in Eq.~\eqref{eq:T}. Every product of the terms $\mathcal{A}$, $\mathcal{B}$, etc.\ in the powers of $\mathcal{L}$ gives rise to a small number of QNPUs; see for example Eq.~\eqref{eq:QNPU}.
The number of circuits to implement the Hamiltonian for the Burgers equation depends on the order of the expansion in $dt$ of $\hat{T}$ in $\hat{C}_2$ and $\hat{T}^\dagger\hat{T}$ in $\hat{C}_1$. Using $\hat{T}$ up to first order, which means $\hat{T}^\dagger\hat{T}$ contains up to $dt^2$, one can implement the Hamiltonian in 18 circuits, as shown in Appendix~\ref{sec:suppl-evaluation}. It is possible to reduce this to 11 circuits by cutting off 7 circuits encoding terms of order $dt^2$ in $\hat{T}^\dagger\hat{T}$. Notably, the number of circuits does not depend on the number of time or space qubits.

The $\hat{C}_0$ term enforcing the initial condition can be measured in a single circuit with a swap test~\cite{buhrman2001quantum} if one is able to prepare $|\psi_0\rangle$ as a quantum state. For this swap test, efficient \textit{destructive} implementations have recently been proposed~\cite{benedetti2021hardware}. While preparing a function in amplitude encoding can require deep circuits in general, most engineering problems have a simple initial state that can be encoded in a shallow circuit on a quantum computer. Alternatively, one could generate this initial state using dissipative engineering~\cite{verstraete2009quantum,horstmann2013noise}.

\subsection{Variational ansatz circuits}\label{sec:method_C}
Throughout this study we parametrize the solution using a variational ansatz circuit
\begin{equation}
    \ket{\Psi} = \hat{U}(\vec{\theta})\ket{0},
\end{equation}
where $\hat{U}(\vec{\theta})$  is a parametrizable quantum circuit of the forms shown in Fig.~\ref{fig2}.
We consider two different parametrizable quantum circuits as variational ansatz to find the ground state of the Hamiltonian $\hat{H}$.
First, we consider a short and dense brickwall ansatz of various depths, of which a single layer can be seen in Fig.~\ref{fig1:bw-circ}. The brickwall ansatz, also referred to as the checkerboard ansatz~\cite{guseynov2023depth}, is especially suited for NISQ devices~\cite{tepaske_optimal_2023}. It has been used before in the context of the Variational Quantum Eigensolver (VQE)~\cite{uvarov2020machine} as well as simulations of quantum dynamics~\cite{mc2023classically}.
Second, we employ a tensor-inspired quantum matrix product state (MPS) ansatz, as seen in Fig.~\ref{fig1:mps-chi-4}.
In classical simulations of dynamical systems, matrix product states represent a promising technique to improve the algorithmic efficiency thanks to their efficient representation of multidimensional problems~\cite{gourianov2022quantum, gourianovDPhil}.
Quantum MPS have been employed successfully for the numerical simulation of one-dimensional systems with relatively small correlations~\cite{haghshenas2022variational}.
Our ansatz consists of successive three-qubit unitaries, which corresponds to a sparse representation of a classical MPS with bond dimension $\chi=4$. By varying the depth of the individual unitaries, we obtain an ansatz with 13, 26 or 39 CNOT gates. This is described in more detail in Appendix~\ref{sec:suppl-ansatz}.

In order to identify an optimal ordering of spacetime qubits that may take advantage of the underlying entanglement structure, we investigate two different ways to order the qubits in the register: a sequential structure with separate time and space registers (blue arrow in Fig.~\ref{fig1:sequential}), and one where the space register is reversed to have the qubits representing the coarsest structures in space and time next to each other (red arrow in Fig.~\ref{fig1:sequential}). This ordering is particularly suitable for the multigrid optimization method discussed below and provides a low-rank tensor representation \cite{lubasch2018multigrid}.

In order to accelerate the minimization procedure and avoid barren plateaus, we initially use small values for coefficients such as the $D$ or $\beta$ in Eq.~\eqref{eq:burgers}. We use the Adam optimizer~\cite{kingma2014adam} to find the ground state of $\hat{H}$ for this simpler problem.
Then, we improve the solution in steps with the L-BFGS-B optimizer~\cite{byrd1995limited}, as described in Appendix \ref{sec:protocols}. In Sec.~\ref{sec:multigrid}, we discuss a multigrid approach to further improve the scalability, limiting the number of variational parameters.

\subsection{Variational optimization on NISQ devices}
\label{sec:method_D}
We want to implement the nonlinear Hamiltonian~\eqref{eq:H} on NISQ devices such as IBMQ Ehningen and Quantinuum H1-1.
IBMQ Ehningen is a superconducting quantum computer with 27 qubits, based on the IBM Quantum Falcon processor which allows for two-qubit gate infidelities below $10^{-2}$ and coherence times around 100 µs~\cite{jurcevic2021demonstration}. 
It features fast computation times, suitable for optimization problems with many iterations.
Quantinuum H1-1 is an ion trap quantum computer with 20 qubits. 
It features all-to-all connectivity, a two-qubit gate infidelity around $2\times10^{-3}$~\cite{quantinuum} and coherence times in the order of seconds~\cite{pino2021demonstration}.
Therefore, complicated circuits are evaluated at high accuracy.

\begin{figure*}
    \centering
    \begin{subfigure}{.49\textwidth}
        \phantomcaption\label{fig2:burgers_quantinuum}
        \phantomcaption\label{fig2:burgers_ibmq}
        \includegraphics[trim={1mm 3mm 0 3mm},clip]{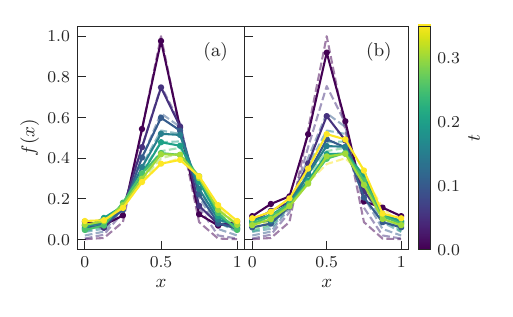}
    \end{subfigure}
    ~
    \begin{subfigure}{.49\textwidth}
        \phantomcaption\label{fig2:diffusion_quantinuum}
        \phantomcaption\label{fig2:diffusion_ibmq}
        \includegraphics[trim={1mm 3mm 0 3mm},clip]{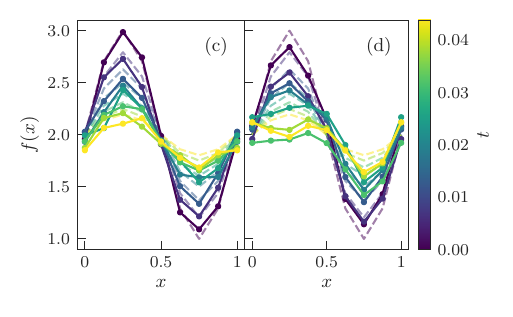}
    \end{subfigure}
    \caption{Solutions to the Burgers and diffusion equation~\eqref{eq:burgers} sampled on NISQ machines. We use $3+3$ qubits ($2^3=8$ points in space and in time) and a reversed space \textit{brickwall ansatz} (Sec.~\ref{sec:method_C}) with 4 layers (Burgers) or 3 layers (diffusion). (a) Burgers equation, Quantinuum H1-1. (b) Burgers equation, IBMQ Ehningen. (c) Diffusion equation, Quantinuum H1-1. (d) Diffusion equation, IBMQ Ehningen.
    The IBMQ Ehningen experiments were run on 17 August 2023, and the Quantinuum H1-1 experiments on 15 and 20 September 2023.
    The dashed lines correspond to a converged circuit from a noiseless simulator. This noiseless result has a cost $\langle\hat{H}\rangle$ of $2.7\times10^{-4}$ (Burgers) or $4.7\times10^{-13}$ (diffusion), and a very good overlap with a numerical solution (infidelity of $3.3\times10^{-4}$ (Burgers) or $3.2\times10^{-7}$ (diffusion).) More details can be found in Sec.~\ref{sec:ibmq-quantinuum}.
    }
    \label{fig2:tomography}
\end{figure*}

After performing the VQE optimization on the Qiskit statevector simulator~\cite{Qiskit}, we sample the resulting state on both quantum computers.
On IBMQ Ehningen, we use readout error mitigation from the M3 library~\cite{nation_scalable_2021}, and transpile the circuits using the highest optimization level. 
This includes dynamical decoupling~\cite{souza2012robust} to avoid decoherence during the idle time of the qubits.  As a result, the two-qubit gate infidelity becomes the main source of errors.
On Quantinuum H1-1, we run our circuit without error mitigation.  The main sources of errors on this machine are qubit depolarization and the two-qubit gate infidelity~\cite{pino2021demonstration}.

We have validated the QNPU implementation for a $2+2$-qubit problem on a noiseless simulator. For further benchmarking of our nonlinear Hamiltonian $\hat{H}$, we decompose it into Pauli strings because this is more feasible for a low number of qubits.
Nonlinear terms such as $f(x,t)\frac{\partial f(x,t)}{\partial x}$ in Eq.~\eqref{eq:burgers} depend on the ansatz state itself.
Therefore, we perform a Pauli decomposition of a linearized Hamiltonian at each iteration of the optimization. 
While this is possible for small systems, the decomposition of an observable into Pauli strings scales exponentially with the number of qubits~\cite{motzoi_linear_2017}.

\section{Results and discussion}
\label{sec:results}

In Sec.~\ref{sec:ibmq-quantinuum}, we perform simulations on the Quantinuum and IBMQ quantum computers to show that our ansatz circuits are viable on NISQ devices. 
In Sec.~\ref{sec:depth}, we investigate the performance of the different ansatz and entanglement structures in the minimization of the Hamiltonian $\hat{H}$ and numerically demonstrate the scaling.
Finally, we discuss promising optimization strategies for our hybrid algorithm in Sec.~\ref{sec:multigrid} and the applicability of our methods beyond variational algorithms in Sec.~\ref{sec:beyond}.

\subsection{IBMQ and Quantinuum}\label{sec:ibmq-quantinuum}

We want to solve the Burgers equation,
\begin{equation}\label{eq:burgers}
    \frac{\partial f(x,t)}{\partial t}=D\frac{\partial^2 f(x,t)}{\partial x^2}-\beta f(x,t)\frac{\partial f(x,t)}{\partial x},
\end{equation}
a nonlinear PDE that describes the convection-diffusion of particles in a viscous medium. Here, 
$D$ is the diffusion coefficient and $\beta$ determines the strength of the nonlinearity.
If one sets $\beta=0$, the nonlinear term vanishes and the diffusion or heat equation is obtained~\cite{fogedby1998solitons}.

First, we study this equation for $3+3$ qubits, which corresponds to a spacetime grid of $2^3=8$ points in time and $2^3=8$ points in space. We use $\beta=1$ and $D=0.05$ with periodic boundary conditions, with a time step of $\Delta t=0.05$. A detailed discussion of the discretization error in spacetime is given in Appendix~\ref{sec:suppl-ibmq}. To emphasize the nonlinearity in the form of a shock wave, we choose a Gaussian profile $f(x,t_0)=\exp(-(2\pi x-\pi)^2)$ for $x\in[0,1]$ as initial state.
On a noiseless statevector simulator, we optimize for the VQE solution $|\psi_{\rm VQE}\rangle$ using a 4-layer brickwall ansatz (circuit depth 8) with reversed space entanglement structure (Fig.~\ref{fig1:sequential}).
The cost $\langle\hat{H}\rangle$ of this state is low, $2.7\times10^{-4}$. The infidelity $1-\langle \psi_{\rm num}| \psi_{\rm VQE} \rangle=3.3\times10^{-4}$, where $|\psi_{\rm num}\rangle$ is the solution from the ODE solver from the SciPy library~\cite{2020SciPy-NMeth}, proves that our VQE optimization has converged. 

We sample this spacetime state $|\psi_{\rm VQE}\rangle$ on the IBMQ Ehningen and Quantinuum H1-1 quantum computers (see Figs.~\ref{fig2:burgers_quantinuum} and \ref{fig2:burgers_ibmq}).
On Quantinuum H1-1, we take $2.5\times10^4$ shots, whereas on IBMQ Ehningen, we take $2\times10^6$ shots (in 20 series of $10^5$ shots), which has a similar execution time.
 We observe that the Quantinuum computer is able to reproduce the rightwards shift of the wave, characteristic for the Burgers equation, whereas the IBM result is more affected by noise, especially in the yellow curve representing the last time point.
The analogous simulation with a quantum MPS ansatz is plotted in Figs.~\ref{fig4:burgers_quantinuum} and \ref{fig4:burgers_ibmq}, where the Quantinuum result stems from an emulator which models the noise of the Quantinuum H1-1 machine.

Next, we discard the nonlinear term from Eq.~\eqref{eq:burgers} by setting $\beta=0$ and $D=1$ with a time step of $\Delta t=0.00625$, which gives us the one-dimensional diffusion or heat equation.
As initial condition at $t_0=0$, we consider a sinusoidal profile $f(x,t_0)=2+\sin(2\pi x)$ for $x\in[0,1]$, with periodic boundary conditions. 
We have chosen the constant $2$ instead of $1$ for the experiments on both quantum computers, because we find that, in the presence of noise, we measure some counts for every bitstring, including the ones that correspond to values close to zero. This makes it difficult to distinguish values close to zero in amplitude encoding.
This shift does not change the dynamics of the diffusion equation.
For comparison, we show the unshifted simulations in Fig.~\ref{fig:suppl-no-shift} in Appendix~\ref{sec:suppl-ibmq}.

\begin{figure}
    \centering
    \begin{subfigure}{\textwidth}
        \includegraphics[trim={1mm 3mm 0 3mm},clip]{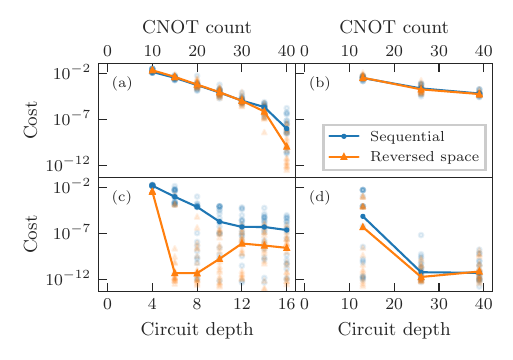}
        \phantomcaption\label{fig4:bw_nonlinear}
        \phantomcaption\label{fig4:mps_nonlinear}
        \phantomcaption\label{fig4:bw_linear}
        \phantomcaption\label{fig4:mps_linear}
    \end{subfigure}
    \caption{Cost $\langle\hat{H}\rangle$ with $\hat{H}$ from Eq.~\eqref{eq:H} as a function of the circuit depth on a noiseless simulator for $3+3$ qubits. (a) Burgers equation, brickwall; (b) Burgers equation, quantum MPS; (c) diffusion, brickwall; (d) diffusion, quantum MPS. The data points in (a) and (c) represent 2 to 8 layers, those in (b) and (d) three quantum MPS ansatzes with 13 to 39 CNOT gates as detailed in Appendix~\ref{sec:suppl-ansatz}. The top and bottom axes indicate that the CNOT count and circuit depth are equivalent for the quantum MPS ansatz but different for brickwall circuits.
    The lines represent the median of the cost resulting from 20 different random initializations, which are plotted as scattered points. The maximum number of iterations is $4\times2,500$ as described in Appendix~\ref{sec:protocols}.}
    \label{fig:depth_6qubits}
\end{figure}

We calculate a VQE solution on the Qiskit statevector simulator using a 3-layer brickwall ansatz (circuit depth 6). The cost $\langle\hat{H}\rangle$ of this state is $4.7\times10^{-13}$.
 To verify that this state with a low cost indeed matches the desired solution, we calculate the infidelity with regards to a solution from numerical integration $1-\langle \psi_{\rm num}| \psi_{\rm VQE} \rangle=3.2\times10^{-7}$. These cost and infidelity values indicate that the upwards shift of the initial condition does not prevent convergence of the VQE. In Figs.~\ref{fig2:diffusion_quantinuum} and \ref{fig2:diffusion_ibmq}, we sample it on the IBMQ Ehningen and Quantinuum H1-1 quantum computers.
The analogous simulation using a quantum MPS ansatz is depicted in Figs.~\ref{fig4:diffusion_quantinuum} and \ref{fig4:diffusion_ibmq}.

\subsection{Circuit depth requirements}\label{sec:depth}

\begin{figure}
    \centering
    \begin{subfigure}[t]{\textwidth}
        \includegraphics[trim={0 3mm 5mm 10mm},clip]{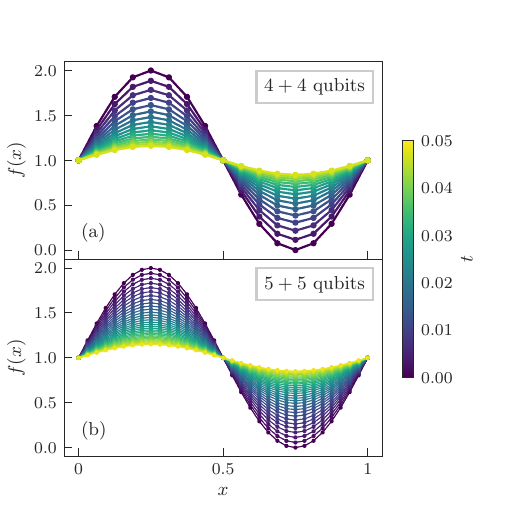}
        \phantomcaption\label{fig:8q}
        \phantomcaption\label{fig:10q}
    \end{subfigure}
    \caption{Results for the diffusion equation (Eq.~\eqref{eq:burgers} with $\beta=0$) obtained from a noiseless simulator for different numbers of qubits.
    (a) Obtained solution for $4+4$ qubits ($2^4=16$ points in space and time) with $D=1$ and $\Delta t=1/320$. Reversed space brickwall ansatz with 4 layers, with a cost of~ $1.6\times10^{-9}$ and an infidelity of~ $9.3\times10^{-8}$. (b) $5+5$ qubits ($2^5=32$ points in space and time) with $D=1$ and $\Delta t=1/640$. Reversed space brickwall ansatz with 6 layers, with a cost of~ $7.0\times10^{-7}$ and an infidelity of~ $2.9\times10^{-7}$.}
    \label{fig:8-10q}
\end{figure}

We show the achieved cost of the VQE solutions for all ansatz and entanglement structures in Fig.~\ref{fig:depth_6qubits}. Note that the initial condition for the diffusion equation is $f(x,t_0)=1+\sin(2\pi x)$ here.
The circuit depth in this figure is defined by counting the consecutive layers of CNOT gates; for the brickwall ansatz with 6 qubits, this is the number of CNOT gates divided by $2.5$ because one brickwall layer contains 5 blocks, but has a depth of 2. For the quantum MPS ansatz, it is equal to the number of CNOT gates.
We find in Fig.~\ref{fig4:bw_linear} that our diffusion problem is very well described using a shallow reversed space ansatz with 3 or 4 brickwall layers. The low cost which we observe for these depths, and the subsequent increase in cost for deeper circuits, indicates that these ansatzes can be optimized in less iterations than the deeper ones.
The sequential ordering (see Fig.~\ref{fig1:sequential}) requires a few more layers and is more sensitive to random initializations. For the nonlinear problem in Fig.~\ref{fig4:bw_nonlinear}, we see that the deeper ansatzes provide an advantage in expressibility, with only small differences between the entanglement structures.
The quantum MPS ansatz requires a slightly higher number of CNOT gates, but is less sensitive to the ordering of the qubits in the register, as seen in Figs.~\ref{fig4:mps_nonlinear} and \ref{fig4:mps_linear}.
We conclude that there are particularly efficient (shallow) representations of our diffusion problem using the reversed space entanglement structure;
 however, the cost for these shallow circuits does show a significant spread depending on the random initial condition, indicating there are local minima to the cost function. This is consistent with literature about the trainability of shallow circuits~\cite{KiKiRo21, Larocca2023}.
Our classical MPS calculations indicate that the entanglement entropy in the ansatz and the necessary ansatz depth depend significantly on the smoothness of the initial condition and the spectral content of its evolution.

\begin{figure}
    \begin{subfigure}[t]{\textwidth}
        \includegraphics[trim={1mm 3mm 0 3mm},clip]{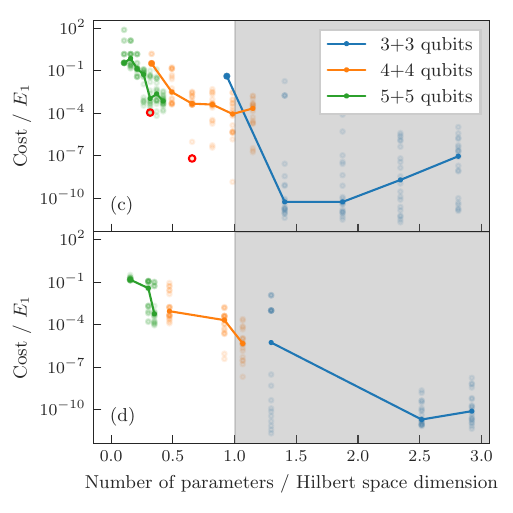}
        \phantomcaption\label{fig:scaling-bw}
        \phantomcaption\label{fig:scaling-mps}
    \end{subfigure}
    \caption{Cost $\langle\hat{H}\rangle$ with $\hat{H}$ from Eq.~\eqref{eq:H} achieved for the diffusion equation with different numbers of qubits, divided by the energy $E_1$ of the first excited state, plotted as a function of number of parameters divided by Hilbert space dimension (64 for 6 qubits, 256 for 8 qubits, 1024 for 10 qubits.) The shaded area indicates  where the number of parameters exceeds the dimension of the Hilbert space. The dots in the background represent the 20 individual runs for which the line shows the median. (a) Brickwall ansatz of 2 to 8 layers.  The two states from Fig.~\ref{fig:8-10q} are marked with red circles. (b) Three quantum MPS circuits, as detailed in Appendix~\ref{sec:suppl-ansatz}. A reversed space entanglement structure is used in both cases. We have increased the number of iterations from $2,500$
    to $5,000$ for every value of $D$ for $8$ qubits, and to $10,000$ for every value of $D$ for $10$ qubits.}
    \label{fig:scaling}
\end{figure}

\begin{figure*}[!t]
    \begin{subfigure}[c]{0.49\textwidth}
        \caption{}\vspace{-2em}
        \begin{quantikz}[thin lines,row sep=0.15cm,column sep=0.4cm,font=\small]
    		\lstick[label style=red]{$t_3$} & \qw & \gate[2,style={draw=red,fill=red!20},label style=red]{U} & \qw & \gate[2,style={draw=red,fill=red!20},label style=red]{U} & \qw \\
            \lstick{$t_2$} & \gate[2,style={draw=blue,fill=blue!20},label style=blue]{U} & \qw & \gate[2,style={draw=blue,fill=blue!20},label style=blue]{U} & \qw & \qw \\
            \lstick{$t_1$} & \qw & \gate[2]{U} & \qw & \gate[2]{U} & \qw \\
            \lstick{$t_0$} & \gate[2]{U} & \qw & \gate[2]{U} & \qw & \qw \\
            \lstick{$x_0$} & \qw & \gate[2]{U} & \qw & \gate[2]{U} & \qw \\
            \lstick{$x_1$} & \gate[2,style={draw=blue,fill=blue!20},label style=blue]{U} & \qw & \gate[2,style={draw=blue,fill=blue!20},label style=blue]{U} & \qw & \qw \\
            \lstick{$x_2$} & \qw & \gate[2,style={draw=red,fill=red!20},label style=red]{U} & \qw & \gate[2,style={draw=red,fill=red!20},label style=red]{U} & \qw \\
            \lstick[label style=red]{$x_3$} & \qw & \qw & \qw & \qw & \qw 
        \end{quantikz}
        \label{fig:multigrid-ansatz}
    \end{subfigure}
    ~
    \begin{subfigure}[c]{0.49\textwidth}
        \caption{}
        \includegraphics[trim={0 3mm 0 3mm},clip]{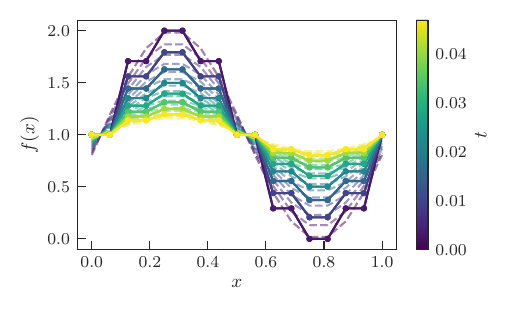}
        \label{fig:multigrid-step}
    \end{subfigure}

    \caption{Multigrid optimization strategy. (a) The reversed space ansatz of two brickwall layers is expanded from 6 to 8 qubits. The red blocks correspond to the new spacetime qubits ($x_3,t_3$) and the optimization is initially only performed for the angles within the red and blue blocks, while the unitaries of the qubits encoding the coarser structures ($x_0,t_0$) are fixed to the parameters of the converged circuit with 6 qubits.
    (b) Converged $3+3$-qubit result expanded to $4+4$ qubits,
    showing the ``step'' profile that is created by initializing the new qubits with a Hadamard gate. The dashed lines correspond to the solution obtained via numerical integration.}
\end{figure*}

We study the scaling of our algorithm. 
To this aim, we analyze the required circuit depth increasing the system size from 6 to 10 qubits ($n_x+n_t=3+3$, $4+4$ and $5+5$). 
This gives us a measure for the necessary coherence time required to run any of these circuits on a real quantum computer.
The results for a 4-layer brickwall ansatz with $4+4$ qubits and a 6-layer brickwall ansatz with $5+5$ qubits are shown in Fig.~\ref{fig:8-10q}. 
The costs for brickwall and quantum MPS ansatzes of different depths are shown in Fig.~\ref{fig:scaling}.  We normalize them by dividing by the energy $E_1$ of the first excited state of the Hamiltonian for the respective number of qubits, such that we can use the distance to the first excited state as a measure of the quality of the solution, irrespective of the number of qubits. We observe in this figure that the low cost solutions for $3+3$ qubits require an ansatz with more parameters than the dimension of the Hilbert space, whereas the solutions from Fig.~\ref{fig:8-10q} don't.
From this analysis, we conclude that a crossover exists between coarse grids requiring an  ''overparametrized'' ansatz for convergence (6 qubits), and finer grids (8 qubits or more) that can be efficiently encoded with less variational parameters than the dimension of the corresponding Hilbert space.

\begin{figure*}
    \begin{subfigure}[c]{0.49\textwidth}
        \caption{}
        \includegraphics[trim={0 3mm 0 3mm},clip]{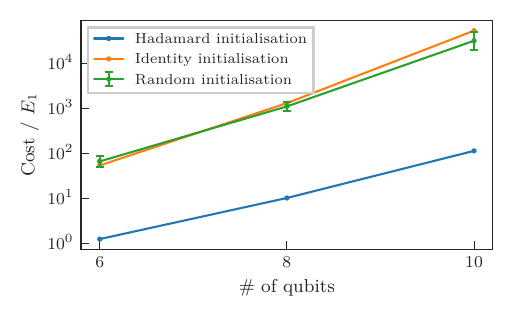}
        \label{fig:multigrid-cost}
    \end{subfigure}
    ~
    \begin{subfigure}[c]{0.49\textwidth}
        \caption{}
        \includegraphics[trim={0 3mm 0 3mm},clip]{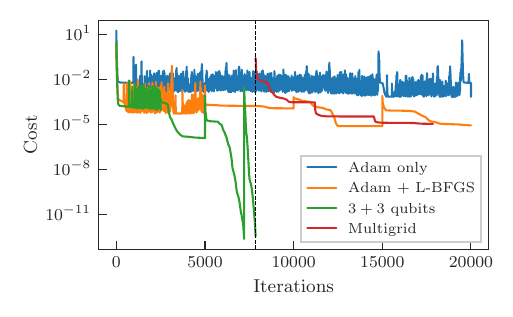}
        \label{fig:multigrid-comparison}
    \end{subfigure}
    \caption{(a) Increase in cost caused by adding more qubits to a converged solution of the diffusion equation, normalized by the energy of the first excited state $E_1$, for different initializations of the new qubits.
    (b) Comparison of the number of iterations versus cost for a $4+4$-qubit problem using different optimization procedures, showing that the multigrid ansatz has a performance comparable to the Adam + L-BFGS procedure.  For all three lines, the best result out of 20 random initializations was chosen. The green line describes an Adam + L-BFGS optimization for $3+3$ qubits, followed by a transition to $4+4$ qubits marked by the dashed black line, and subsequent L-BFGS optimization  in red. This illustrates our multigrid strategy. The  green and red lines for the multigrid strategy are shorter because the $3+3$-qubit simulation considered only took 7,857 iterations instead of the total 10,000.}
    \label{fig:multigrid}
\end{figure*}

\subsection{Barren plateaus and multigrid optimization}\label{sec:multigrid}

Variational cost functions commonly exhibit barren plateaus: gradients of the cost function exponentially vanish as the number of qubits increases when optimizing using random initial parameters~\cite{mcclean2018barren}.
Fig.~\ref{fig4:bw_linear} indicates that it can be beneficial to optimize for a smaller depth first and then increase the depth if necessary.
This motivates our approach of first solving a simpler problem with a small $D$ and/or $\beta$ and then ramping up these physical constants (see Appendix \ref{sec:protocols}).
We calculate the variance of the gradient for different random initializations in Appendix~\ref{sec:suppl-barren-plateau} to evaluate this strategy. We do not observe any barren plateaus for the system sizes studied here.

Now, we analyze a multigrid approach~\cite{lubasch2018multigrid} for further improvements: first, the solution is found for a coarse grid using a combination of the Adam and L-BFGS optimizers, as described in Appendix~\ref{sec:protocols}; second, optimization on a finer grid is started with the optimal parameters of the coarse grid. 
We use the reversed space entanglement structure. In this way, it becomes possible to add new time or space qubits while maintaining the ansatz structure for the coarser grid, as shown in Fig.~\ref{fig:multigrid-ansatz}. 
This strategy is inspired by the fact that the entanglement entropy is greatest for the tensors encoding the coarsest partitions of the grid in a matrix product state representation \cite{lubasch2018multigrid}.

In order to stay close to the ground state and avoid detrimental jumps in the cost when transitioning from a coarse to a fine grid, 
we want to implement a `step' profile as shown in Fig.~\ref{fig:multigrid-step}.
We achieve this by slightly modifying the brickwall ansatz so that every 2-qubit unitary contains two CNOT gates instead of one, and six rotations after every CNOT, duplicating the unit seen in Fig.~\ref{figSM:bw_2qu} in Appendix~\ref{sec:suppl-ansatz}. 
There, we also show that this modification of the ansatz does not negatively affect the required circuit depth, because less layers are required if every layer contains twice the number of gates.
By initializing all of these angles to zero, except for the very last rotations on the new qubits, the new qubits representing fine structures in space and time are not immediately entangled with the existing coarse ones. 
Importantly, the rotations at the end of the ansatz are initialized such that they represent a Hadamard gate. This results in a `step' profile after initialization of the fine grid ansatz, as shown in Fig.~\ref{fig:multigrid-step}. 
We slightly shift the initial condition along the $x$-axis to make the generated step profile a better match. 
In Fig.~\ref{fig:multigrid-cost}, we depict the cost after transitioning from a coarse to a fine grid. Our `step' profile algorithm results in a smaller cost function than when initializing the new qubits randomly or in a zero state. It may be possible to engineer other initial states, such that they come even closer to the desired solution.

After adding the new qubits, we optimize for the new parameters and the blocks affecting the qubits $x_2$ and $t_2$ directly adjacent to the new ones, as shown in Fig.~\ref{fig:multigrid-ansatz}. 
This guides the optimizer towards the solution of the refined problem. 
After this optimization round, we add the blocks affecting $x_1$ and $t_1$ and optimize again.
Finally, we optimize all blocks.
We compare the cost as a function of the number of iterations of this method with a direct optimization for $4+4$ qubits in Fig.~\ref{fig:multigrid-comparison}, either using Adam only, without ramping up $D$ on the way, or using Adam and L-BFGS. This figure shows that the multigrid strategy, with a transition from $3+3$ to $4+4$ qubits, needs a similar number of iterations as a direct optimization for $4+4$ qubits with Adam and L-BFGS, whereby the multigrid method is faster because it optimizes less parameters. It is also clear that either method provides a significant advantage compared to solving with Adam only, which ends up in a local minimum almost two orders of magnitude higher than the other methods in this example. We conclude that the multigrid strategy described here is a promising method to solve problems on fine grids.

\subsection{Beyond variational algorithms}\label{sec:beyond}
The nonlinear Feynman--Kitaev Hamiltonian is applicable beyond the variational methods discussed in this article. Besides the variational quantum eigensolver (VQE), on which our methods are based, there are many other variational and non-variational techniques for finding the ground state. Imaginary time evolution, which we used in Appendix~\ref{sec:uniqueness} to study the stability of the solution, has a quantum analog that is applicable on NISQ devices~\cite{motta2020determining}. This quantum imaginary time evolution, which has previously been used to solve differential equations~\cite{alghassi2022variational}, is still variational.
A non-variational strategy can be found in adiabatic quantum computing~\cite{albash2018adiabatic}, which has similarities with the optimization strategy we used of increasing $D$ and/or $\beta$ in steps. An implementation on NISQ devices has been recently shown~\cite{hegade2021shortcuts} and promising proposals exist to further compress the corresponding quantum circuits~\cite{mc_keever_towards_2024}. As such, we expect that this can also be used to solve our ground-state problem.

\section{Conclusions}\label{sec:conclusion}

In this work, we present an extension of the Feynman–Kitaev formalism,  originally developed for quantum dynamics, that is tailored to the integration of arbitrary PDEs with nonlinearities. 
We provide proof-of-principle calculations that demonstrate that nonlinear dissipative processes are well reproduced within this framework. 
Here, the full spacetime solution of the system can be retrieved in a single optimization routine of an appropriate cost function, which prevents the accumulation of errors of iterative time marching schemes.
We see this as a necessary step to solve PDEs on quantum computers, as stability conditions (see Appendix~\ref{sec:uniqueness} for details) require that an increase in spatial resolution goes together with an increase in time resolution.
Our results indicate that solutions to nonlinear PDEs can be obtained with shallow circuits using either a brickwall ansatz with a few layers or a sparse formulation of matrix product states (MPS) as a quantum circuit.
We adopt a multigrid strategy where the ansatz is adaptively optimized for finer and finer solutions of the discretized PDE, thereby making the spacetime quantum solver potentially scalable in the number of qubits.
We show how to adapt the ansatz so it can be expanded with additional qubits while keeping the coarser result. We expect this multigrid strategy to become more powerful as the resolution, and thereby smoothness of the function, increases.
We adapt the Quantum Nonlinear Processing Unit (QNPU) \cite{lubasch2020variational} to evaluate the cost function.  To implement time evolution, we show how to implement derivatives without periodic boundary conditions. Notably, the number of circuits to be measured is only dependent on the number of (non)linear terms in the PDE and the desired accuracy of the time evolution operator and not on the number of qubits, and the depth of the circuits is linear in the number of qubits. 
Depending on the nature of the problem and the available quantum resources, it can thus be favorable to use a first or second order propagator and a fine grid in time,  instead of a higher order integration scheme on a coarser time grid, which is commonly done in classical simulations~\cite{li_numerical_2017}. Alternatively, one could start with a low order solution and use that as input for a higher order optimization with more circuits.

Current classical methods for nonlinear PDEs in fluid dynamics allow around $10^6$ spatial points~\cite{gourianov2022quantum,mani2023perspective}. This amounts to about $10^2$ points in every dimension for a 3D grid, and is equivalent to 20 spatial qubits. Hence we believe that only around 100 qubits in total (including time qubits and duplicates of the ansatz for nonlinearities) would be necessary to go beyond current classical simulations.
Our methods are also relevant in classical computing, as recent developments in tensor network representations of nonlinear problems have shown that it is possible to devise efficient classical algorithms inspired by quantum algorithms for PDEs~\cite{gourianov2022quantum,gourianovDPhil,kiffner_tensor_2023}.
Finally, we remark that our methodology does not require a variational approach for finding the ground state of the Hamiltonian, and alternative techniques can be readily employed.

\begin{acknowledgments}
This project was made possible through the support from the Competence Center Quantum Computing Baden-Württemberg in the context of the QuESt and QuESt+ projects (funded by the Ministry of Economic Affairs, Labour and Tourism Baden-Württemberg). 
The classical computations were performed on the JUSTUS 2 cluster, supported by the state of Baden-Württemberg through bwHPC and the German Research Foundation (DFG) through grant no. INST 40/575-1 FUGG.
We are grateful to Eric Brunner and Chris N. Self for proofreading the manuscript and valuable discussions.
\end{acknowledgments}

\appendix

\section{Derivation of the cost function}\label{sec:derivation}

We start with the general form of the differential equation as given by Eq.~\eqref{eq:diffeq},
where the differential operator ${\mathcal{L}}$ (linear or non-linear) is the generator of the underlying dynamics. For simplicity, let us assume that ${\mathcal{L}}$ is time-independent. We encode the values of the function in a quantum state using the amplitude encoding of Eq.~\eqref{eq:amplitude}. This way we can define a quantum version of Eq.~\eqref{eq:diffeq} as
\begin{equation}\label{eq:quantum-diff-eq}
    \frac{d}{dt}\ket{\Psi}=\mathcal{L}\ket{\Psi},
\end{equation}
where the amplitudes $\psi(x,t)=\mathcal{N}f(x,t)$ of the state $\ket{\Psi}$ represent the function values up to a certain normalization constant $\mathcal{N}$.
A general solution to this equation can be written in terms of the time evolution operator (propagator),
\begin{equation}
    \vec{\psi}(t) = \hat{T}(t)\vec{\psi}_0,
\end{equation}
where $\vec{\psi}(t)$ is the state in space for a specific time $t$.
We use this propagator to define an implicit integration scheme
\begin{equation}\label{eq:suppl-implicit}
dt(\partial_t - \mathcal{L})\vec{\psi}(t_i) \approx \hat{T}(-dt) \vec{\psi}(t_{i+1}) - \vec{\psi}(t_i),
\end{equation}
where the vector notation $\vec{\psi}(t_i)$ is used for the state in space at time step $\ket{i}$ such that $\vec{\psi}(t_i)=\braket{i}{\Psi}$. For a time-dependent $\mathcal{L}$, one would replace the backwards propagator $\hat{T}(-dt)$ by its time-dependent equivalent $\hat{T}(t+dt,t)$.
If $\hat{T}$ is approximated to first order in $dt$, this is equivalent to a backward Euler integration scheme. We turn this into a spacetime representation by multiplying on the left with $\ket{i}$ and taking the sum over all $i$, which leads to
\begin{align}
    dt(\partial_t-\mathcal{L})|\Psi\rangle &\approx \! \sum_{i = 0}^{N_{t}-1} \! \left(\hat{T}(-dt)\otimes|i\rangle \langle i+1| - \hat{I}\otimes|i\rangle \langle i|\right)\!|\Psi\rangle \nonumber \\
    &\equiv\hat{X}\ket{\Psi},
\end{align}
where $\hat{X}$ is as defined in Eq.~\eqref{eq:X}.
This means that $\hat{X}\approx dt(\partial_t-\mathcal{L})$. 
We multiply with the adjoint to get an Hermitian operator $\hat{X}^\dagger\hat{X}\approx dt^2\|\partial_t-\mathcal{L}\|^2$, and use this operator to define a ground-state problem for Eq.~\eqref{eq:diffeq}. Combined with the $\hat{C}_0$ term for the initial condition $\ket{\psi(t=0)}=\ket{\psi_0}$, one obtains the Hamiltonian in Eq.~\eqref{eq:H}.

For a better physical understanding, we rewrite $\hat{X}^\dagger\hat{X}$ as
\begin{widetext}
\begin{subequations}
\begin{align}
 &\hat{X}^\dagger\hat{X}= \left( \sum_{j = 0}^{N_{t}-1} \hat{T}^{\dag}(-dt) \otimes |j+1\rangle \langle j| - \hat{I}\otimes |j\rangle \langle j| \right)
 \left( \sum_{i = 0}^{N_{t}-1} \hat{T}(-dt) \otimes |i\rangle \langle i+1| - \hat{I}\otimes |i\rangle \langle i| \right),\\
 &=\sum_{i = 0}^{N_{t}-1} \left( \hat{I}\otimes |i\rangle \langle i| + \hat{T}^{\dag}(-dt)\hat{T}(-dt)\otimes |i+1\rangle \langle i+1| \right)
 - \sum_{i = 0}^{N_{t}-1} \left( \hat{T}^{\dag}(-dt) \otimes |i+1\rangle \langle i| + \hat{T}(-dt) \otimes |i\rangle \langle i+1| \right). \label{eq:cost-b}
\end{align}
\end{subequations}
\end{widetext}
We define the two terms of equation~\eqref{eq:cost-b} as $\hat{C}_1$ and $\hat{C}_2$. We can then alternatively write Eq.~\eqref{eq:H} as
\begin{equation}\label{eq:suppl-H}
    \hat{H}=c_0\hat{C}_0 + \hat{C}_1 - \hat{C}_2,
\end{equation}
with the three terms given by
\begin{subequations}\label{eq:suppl-H-terms}
    \begin{align}
        \hat{C}_0 &= \left[\hat{I}-|\psi_0\rangle\langle\psi_0|\right]\otimes|0\rangle\langle 0|, \label{eq:suppl-c0} \\
        \hat{C}_1 &= \sum_{i = 0}^{N_{t}-1} \left[\hat{I}\otimes |i\rangle \langle i| + \hat{T}^{\dag}(-dt)\hat{T}(-dt)\otimes |i+1\rangle \langle i+1|\right], \label{eq:suppl-c1} \\
        \hat{C}_2 &= \sum_{i = 0}^{N_{t}-1} \left[\hat{T}^{\dag}(-dt) \otimes |i+1\rangle \langle i| + \hat{T}(-dt) \otimes |i\rangle \langle i+1|\right]. \label{eq:suppl-c2}
    \end{align}
\end{subequations}
Here $\hat{C}_0$ ensures the initial condition is kept, $\hat{C}_1$ ensures that all time points are present in the solution, and $\hat{C}_2$ ensures that the steps between them correspond to the application of the differential operator $\mathcal{L}$. For quantum systems, it is possible to derive this cost function from a time-embedded discrete variational principle (TEDVP) that is equivalent to the Dirac--Frenkel--McLachlan time-dependent variational principle in the limit of infinitesimal time \cite{jarrod_mcclean_feynmans_2013}.

\begin{figure*}
\centering
    \begin{subfigure}[c]{0.65\textwidth}
        \caption{}
        \includegraphics[trim={1mm 3mm 0 3mm},clip]{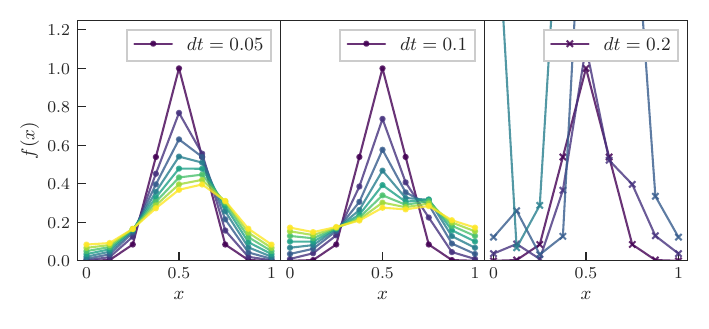}
        \label{fig:9a}
    \end{subfigure}
    ~
    \begin{subfigure}[c]{0.33\textwidth}
        \caption{}
        \includegraphics[trim={0 3mm 1mm 3mm},clip]{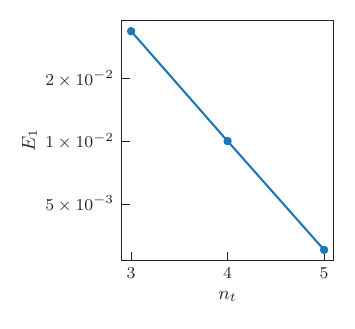}
        \label{fig:9b}
    \end{subfigure}
    \caption{(a) Ground-state profiles of $\hat{H}$ for the Burgers equation ($D=0.05$, $\beta=1$) calculated with imaginary time evolution (ITE).
    Converged solutions can be achieved for $dt=0.05$ and $dt=0.1$, but the breakdown of stable solutions to the discretized PDE with $dt=0.2$ results in unphysical states.
    (b) Energies of the first excited state $E_1$~\eqref{eq:Psi1_eigcondition} of $\hat{H}$ as a function of the number of time qubits $n_t$ for a fixed time step of $dt=0.05$ and a spatial register of $n_x=3$ qubits.}\label{fig:suppl-stability}
\end{figure*}

We show here that the history state~\eqref{eq:history} is an eigenstate of this Hamiltonian with an eigenvalue of 0. To this end, we calculate the action of all three terms of Eq.~\eqref{eq:suppl-H-terms} on this state. The first term gives us
\begin{align}
    \hat{C}_0\ket{\Psi}=&\frac{1}{\sqrt{N_t+1}}\left[\hat{I}-\ket{\psi_0}\bra{\psi_0}\right]\ket{\psi_0}\otimes\ket{0} \nonumber \\
    =& 0\ket{\Psi}.
\end{align}
The second and third terms give us
\begin{subequations}
    \begin{align}
        \hat{C}_1\ket{\Psi}=&\frac{1}{\sqrt{N_t+1}}\sum_{i = 0}^{N_t-1} \left[ \hat{T}^{\,i}(dt)\ket{\psi_0}\otimes\ket{i} \right. \\
        &+ \left. \hat{T}^\dagger(-dt)\hat{T}(-dt)\hat{T}^{\,i+1}(dt)\ket{\psi_0}\otimes\ket{i+1} \right], \nonumber \displaybreak[1] \\
        \hat{C}_2\ket{\Psi}=&\frac{1}{\sqrt{N_t+1}}\sum_{i = 0}^{N_{t}-1} \left[ \hat{T}(-dt) \hat{T}^{\,i+1}(dt)\ket{\psi_0}\otimes\ket{i} \right. \nonumber \\
        &+ \left. \hat{T}^\dagger(-dt)\hat{T}^{\,i}(dt)\ket{\psi_0}\otimes\ket{i+1} \right].
    \end{align}
\end{subequations}
Using the property $\hat{T}(-dt)\hat{T}(dt)=\hat{I}$, one can see that $\hat{C}_1\ket{\Psi}=\hat{C}_2\ket{\Psi}$. Therefore
\begin{equation}
    \hat{H}\ket{\Psi}=\left(c_0\hat{C}_0+\hat{C}_1-\hat{C}_2\right)\ket{\Psi}=0\ket{\Psi},
\end{equation}
which means that $\ket{\Psi}$ is an eigenstate of $\hat{H}$ with eigenvalue 0. We numerically show in Appendix~\ref{sec:uniqueness} that this is the ground state, and that there is a gap with the first eigenstate which makes this ground state unique.

\subsection{Note on implicit integration}

Using the implicit integration scheme of Eq.~\eqref{eq:suppl-implicit}, the $\hat{C}_1$ term~\eqref{eq:suppl-c1} contains a term with $\hat{T}^{\dag}(-dt)\hat{T}(-dt)$. If one performs the derivation with an explicit integration scheme, there is a term $\hat{T}^\dag(dt)\hat{T}(dt)$ instead. For quantum dynamics~\cite{jarrod_mcclean_feynmans_2013,barison2022variational}, the propagator is unitary which means that both terms become an identity. In this case, there is no difference between an implicit and an explicit integration scheme. For our nonunitary $\hat{T}$, we have chosen to use an implicit scheme.

\subsection{Uniqueness and stability}\label{sec:uniqueness}
In order to produce stable solutions in a finite-difference integration scheme, the distance travelled by the solution in one time step must be less than the distance between two points in the grid~\cite{li_numerical_2017}.
For the Burgers equation~\eqref{eq:burgers}, this stability condition translates into the requirement that $D\,dt/dx^2 <1/2$. With our model parameters ($D=0.05$, $\beta=1$), we expect stable solutions for $dt\lesssim 0.15$ for 3 space qubits ($dx=1/8$).
To obtain stable solutions from the ground-state problem given by the Feynman--Kitaev Hamiltonian $\hat{H}$, it is necessary that the ground state of $\hat{H}$ is unique. This is ensured if there is an energy gap between the ground state and the first excited state. In this section, we show numerical simulations with imaginary time evolution (ITE) using the \texttt{mpnum} library~\cite{mpnum} to evaluate the existence of this gap under different model parameters.

The ground and first excited states of $\hat{H}$ must satisfy the conditions
\begin{align}
\hat{H}(\ket{\Psi_0})\ket{\Psi_0} &=E_0 \ket{\Psi_0}, \label{eq:Psi0_eigcondition}\\
\hat{H}(\ket{\Psi_1})\ket{\Psi_1} &=E_1 \ket{\Psi_1}. \label{eq:Psi1_eigcondition}
\end{align}
Due to the non-linear nature of the Hamiltonian, the states $\ket{\Psi_0}$, $\ket{\Psi_1}$ may not be orthogonal to each other. Furthermore, the lowest eigenvector $\ket{E'_0}$ of $\hat{H}(\ket{\phi})$ for an arbitrary state $\ket{\phi}$ may not coincide with $\ket{\Psi_0}$.

The ITE method is based on the fact that the operator $e^{-t\hat{H}}$ becomes a projector onto the ground state of $\hat{H}$ in the limit $t\rightarrow \infty$. 
For an arbitrary, initial state $\ket{\phi}$ with a non-negligible overlap $|\langle \Psi_0|\phi\rangle|>0$ with the ground state, the ITE operator $e^{-t\hat{H}}$ acting on $\ket{\phi}$ converges to the ground state
\begin{align}
   \ket{\Psi_0} \propto \lim_{t\rightarrow \infty} \left( e^{-t \hat{H}}\right) \ket{\phi}.
\end{align}
Starting from a random state $\ket{\phi}$, we choose an Euler integration scheme for ITE by iterating
\begin{enumerate}
\item $|\phi'(t+\tau)\rangle = |\phi(t)\rangle -\tau \hat{H} |\phi(t)\rangle$,
\item $|\phi(t+\tau)\rangle = |\phi'(t+\tau)\rangle / \sqrt{\langle\phi'(t+\tau)|\phi'(t+\tau)\rangle}$,
\end{enumerate}
until convergence is reached for $\hat{H}(\ket{\Psi_0})\ket{\Psi_0}=E_0 \ket{\Psi_0}$ with $E_0 \lesssim 10^{-14}$. We normalize the state after each time step, because the norm $\langle\phi(t)|\phi(t)\rangle$ diverges exponentially with $t$.

We turn the calculation of the first excited state $\ket{\Psi_1}$~\eqref{eq:Psi1_eigcondition} into another ground-state problem. This can be achieved by performing ITE with a modified Hamiltonian $\hat{H}^\prime$, whose ground state energy is raised above the energy of its first excited state. One may be tempted to use the ``shifted" Hamiltonian 
\begin{equation}
    \hat{H}^\prime = \hat{H} + c \ket{\Psi_0}\bra{\Psi_0}, \label{eq:shift_nonorthogonal}
\end{equation}
\noindent where $c$ is a positive constant and $\ket{\Psi_0}$ is the ground state  calculated with ITE as described above. However, for non-linear Hamiltonians, the ground and excited states as defined in Eqs.~\eqref{eq:Psi0_eigcondition}, \eqref{eq:Psi1_eigcondition} may not be orthogonal ($\langle \Psi_0|\Psi_1\rangle \neq 0$), and the shift $c \ket{\Psi_0}\bra{\Psi_0}$ in Eq.~\eqref{eq:shift_nonorthogonal} not only modifies the lowest energy subspace of $\hat{H}(\ket{\phi})$, but it also perturbs the rest of the spectrum. To avoid this, we perform ITE simulations with the modified Hamiltonian
\begin{equation}
    \hat{H}^\prime(\ket{\phi}) = \hat{H}(\ket{\phi}) + c \ket{E'_0}\bra{E'_0}, \label{eq:adapt_ITE}
\end{equation}
where the second term raises the energy of the lowest eigenstate $\ket{E'_0}$ of $\hat{H}(\ket{\phi(t)})$, which changes with every ITE iteration. This way the ITE is directed towards a state that solves Eq.~\eqref{eq:Psi1_eigcondition}. The only requisite for the constant $c$ is that it should be larger than the second eigenvalue $E'_1$ of $\hat{H}(\ket{\phi})$ for any $\ket{\phi}$.
We use this method to calculate the energy gap of the nonlinear Hamiltonian $\hat{H}$ associated with the Burgers equation~\eqref{eq:burgers}.

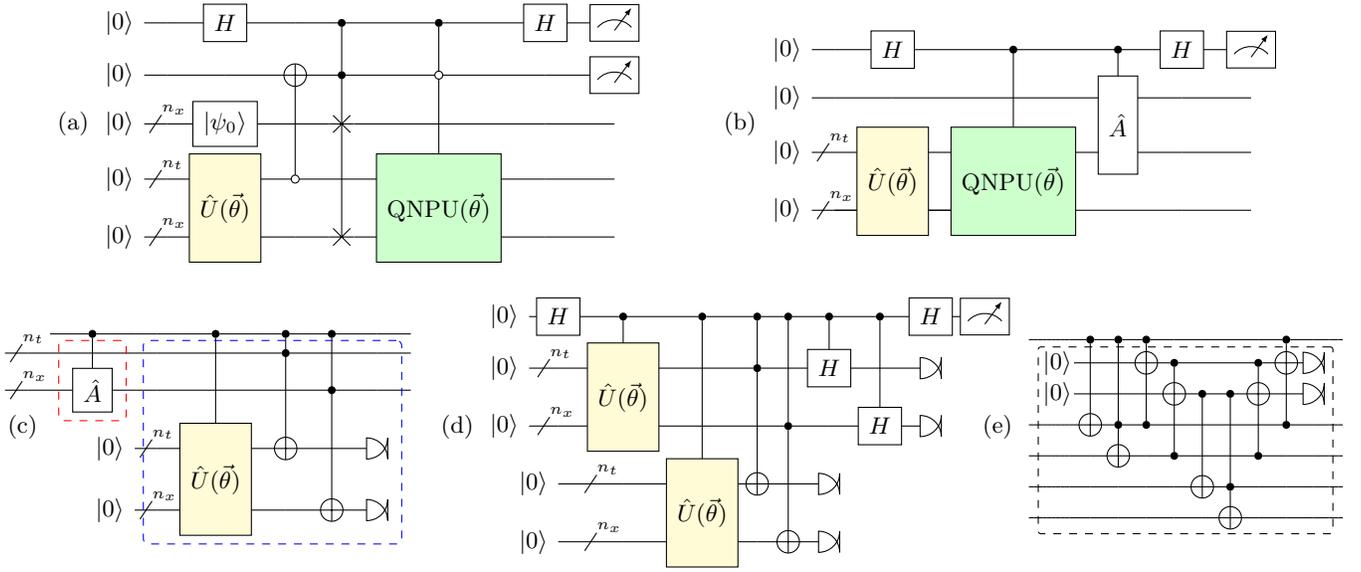
\begin{figure*}
    \begin{subfigure}{0.46\textwidth}
        \centering
        {\small (a)}
        \begin{quantikz}[thin lines,font=\small,row sep=0.1cm,column sep=0.3cm]
            \lstick{\ket{0}} & \qw & \gate{H} & \qw & \ctrl{1} & \ctrl{1} & \gate{H} & \meter{} \\
            \lstick{\ket{0}} & \qw & \qw & \targ{} & \control{} & \octrl{2} & \qw & \meter{} \\
            \lstick{\ket{0}} & \qwbundle{n_x} & \gate{\ket{\psi_0}} & \qw & \targX{}  & \qw & \qw & \qw \\
            \lstick{\ket{0}} & \qwbundle{n_t} & \gate[2,style={fill=yellow!20}]{\hat{U}(\vec{\theta})} & \octrl{-2} & \qw & \gate[2,style={fill=green!20}]{\textrm{QNPU}(\vec{\theta})} & \qw & \qw \\
            \lstick{\ket{0}} & \qwbundle{n_x} & & \qw & \swap{-3} &  & \qw & \qw
        \end{quantikz}
         \phantomcaption\label{fig:suppl-circ-c0-c1}
    \end{subfigure}
    ~
    \begin{subfigure}{0.47\textwidth}
    \centering
    {\small (b)}
    \begin{quantikz}[thin lines,font=\small,row sep=0.1cm,column sep=0.3cm]
        \lstick{\ket{0}} & \qw & \gate{H} & \ctrl{2} & \ctrl{1} & \gate{H} & \meter{}\\
        \lstick{\ket{0}} & \qw & \qw & \qw & \gate[2]{\hat{A}} & \qw & \qw \\
        \lstick{\ket{0}} & \qwbundle{n_t} & \gate[2,style={fill=yellow!20}]{\hat{U}(\vec{\theta})} & \gate[2,style={fill=green!20}]{\textrm{QNPU}(\vec{\theta})} & & \qw & \qw \\
        \lstick{\ket{0}} & \qwbundle{n_x} & \qw & \qw & \qw & \qw & \qw
    \end{quantikz}
    \phantomcaption\label{fig:suppl-circ-c2}
    \end{subfigure}
    
    \begin{subfigure}{0.3\textwidth}
        \centering
        \vspace{2em}
        {\small (c)\hspace{-2em}}
        \begin{quantikz}[thin lines,font=\small,row sep=0.15cm,column sep=0.3cm]
            & & & \ctrl{2} & \qw & \qw & \ctrl{3} & \ctrl{3} & \ctrl{4} & \qw & \qw \\
            \qw & \qwbundle{n_t} & \qw
              & \qw \gategroup[2,steps=1,style={dashed,rounded corners=2pt,red, inner xsep=2pt, inner ysep=0}]{} & \qw
              & \qw \gategroup[4,steps=5,style={dashed,rounded corners=2pt,blue, inner xsep=2pt, inner ysep=0}]{}
              & \qw & \control{} & \qw & \qw & \qw \\
            \qw & \qwbundle{n_x} & \qw & \gate{\hat{A}} & \qw & \qw & \qw & \qw & \control{} & \qw & \qw \\
            & & & & \lstick{\ket{0}} & \qwbundle{n_t} & \gate[2,style={fill=yellow!20}]{\hat{U}(\vec{\theta})} & \targ{} & \qw & \meterD{} \\
            & & & & \lstick{\ket{0}} & \qwbundle{n_x} & & \qw & \targ{} & \meterD{} 
        \end{quantikz}
         \phantomcaption \label{fig:suppl-qnpu_a}
    \end{subfigure}
    ~
    \begin{subfigure}{0.43\textwidth}
        \centering
        {\small (d)}
        \begin{quantikz}[thin lines,font=\small,row sep=0.1cm,column sep=0.1cm]
            \lstick{\ket{0}} & \gate{H} & \ctrl{1} & \ctrl{3} & \ctrl{1} & \ctrl{2} & \ctrl{1} & \ctrl{2} & \gate{H} & \meter{} \\
            \lstick{\ket{0}} & \qwbundle{n_t} & \gate[2,style={fill=yellow!20}]{\hat{U}(\vec{\theta})} & \qw & \ctrl{2} & \qw & \gate{H} & \qw & \meterD{} \\
            \lstick{\ket{0}} & \qwbundle{n_x} & & \qw & \qw & \ctrl{2} & \qw & \gate{H} & \meterD{} \\
            & \lstick{\ket{0}} & \qwbundle{n_t} & \gate[2,style={fill=yellow!20}]{\hat{U}(\vec{\theta})} & \targ{} & \qw & \meterD{} \\
            & \lstick{\ket{0}} & \qwbundle{n_x} &  & \qw & \targ{} & \meterD{} \\
        \end{quantikz}
         \phantomcaption \label{fig:suppl-qnpu_psi2}
    \end{subfigure}\hspace{-2.5em}
    ~
    \begin{subfigure}{0.28\textwidth}
        \centering
        \phantomcaption\label{fig:suppl-adder}
        {\small (e)}
        \begin{quantikz}[thin lines,font=\small,row sep=0.1cm,column sep=0.6mm]
            \qw & \qw & \qw & \qw & \qw & \qw & \qw & \qw & \qw & \qw & \qw & \ctrl{3} & \ctrl{3} & \ctrl{1} & \qw & \qw & \qw & \qw & \ctrl{1} & \qw & \qw & \qw & \qw & \qw \\
            & & & & \gategroup[6,steps=16,style={dashed,rounded corners=1pt,inner ysep=-1.5pt,inner xsep=0}]{} & & & & & & \lstick[label style={inner xsep=0}]{\ket{0}} & \qw & \qw & \targ{} & \ctrl{1} & \qw & \qw & \ctrl{1} & \targ{} & \meterD{} & & & & \\
            & & & & & & & & & & \lstick[label style={inner xsep=0}]{\ket{0}} & \qw & \qw & \qw & \targ{} & \ctrl{3} & \ctrl{3} & \targ{} & \qw & \meterD{} & & & & \\
            \qw & \qw & \qw & \qw & \qw & \qw & \qw & \qw & \qw & \qw & \qw & \targ{} & \ctrl{1} & \ctrl{-2} & \qw & \qw & \qw & \qw & \ctrl{-2} & \qw & \qw & \qw & \qw & \qw \\
            \qw & \qw & \qw & \qw & \qw & \qw & \qw & \qw & \qw & \qw & \qw & \qw & \targ{} & \qw & \ctrl{-2} & \qw & \qw & \ctrl{-2} & \qw & \qw & \qw & \qw & \qw & \qw \\
            \qw & \qw & \qw & \qw & \qw & \qw & \qw & \qw & \qw & \qw & \qw & \qw & \qw & \qw & \qw & \targ{} & \ctrl{1} & \qw & \qw & \qw & \qw & \qw & \qw & \qw \\
            \qw & \qw & \qw & \qw & \qw & \qw & \qw & \qw & \qw & \qw & \qw & \qw & \qw & \qw & \qw & \qw & \targ{} & \qw & \qw & \qw & \qw & \qw & \qw & \qw
        \end{quantikz}
    \end{subfigure}
        \caption{(a) Circuit to evaluate $\hat{C}_0$ and the $\hat{T}^\dagger\hat{T}$ term in $\hat{C}_1$ simultaneously, using an ancilla qubit to determine whether to apply a swap test for $\hat{C}_0$ if the time is $\ket{0}$, or a QNPU for $\hat{C}_1$ otherwise. (b) Circuit to evaluate $\hat{C}_2$, with an additional qubit for the time shift operator to make it non-periodic. This qubit represents the most significant bit of the adder. In these two figures, the notation $\text{QNPU}(\vec{\theta})$ is used as a short-hand for the QNPU itself and any duplicates of the ansatz, which depends on $\vec{\theta}$, encoded on ancilla qubits. (c)~Building blocks of QNPU circuits, with the $\hat{A}$ operator red and the pointwise multiplication $\hat{c}$ in blue. The multi-controlled gates need to be understood as to be applied pairwise: qubit 0 on one time or space register controls qubit 0 on the other register, and so forth for every time and space qubit. (d)~Circuit calculating $\Re(\sum_i\psi_i^2)/\sqrt{N_tN_x}$ with $\Re$ the real value. (e)~Adder circuit for 4 qubits~\cite{lubasch2020variational}.}\label{fig:suppl-circuits}
\end{figure*}

In Fig.~\ref{fig:9a}, we calculate the ground state for the Burgers equation~\eqref{eq:burgers} using the same parameters as in the main text ($D=0.05$, $\beta=1$), for a spacetime grid with $3+3$ qubits ($n_x=n_t=3$) and three different time steps ($dt=0.05,~0.1,~0.2$).
Our results demonstrate a stable solution for the cases with $dt=0.05$ and $dt=0.1$. However, for $dt=0.2$, the ITE trajectories converge to an unphysical state, which is in accordance with the stability condition. In this scenario, the coefficients $\psi_{ij}$~\eqref{eq:amplitude} corresponding to later times become overrepresented.
In Fig.~\ref{fig:9b}, we show $E_1$ as a function of the number of time qubits, for a fixed time step of $dt=0.05$ and a spatial register of $n_x=3$ qubits.
We see that for this time step, there is a gap for any number of time qubits, and that it decreases with the number of qubits for a constant $dt$ as the number of eigenstates of the system increases. 
Our simulations with the modified Hamiltonian~\eqref{eq:adapt_ITE} for $dt=0.2$ show a constantly decreasing energy, without converging to a state that satisfies the eigenvalue condition~\eqref{eq:Psi1_eigcondition}. For the states $\ket{\xi}$ which we found on the way, $\hat{H}(\ket{\xi})$ presents a degenerate ground state. This suggests that there is indeed a closing gap for our modified Feynman--Kitaev Hamiltonian once we violate the stability condition.

\section{Evaluation of the cost function}\label{sec:suppl-evaluation}

In order to measure the expectation value $\langle\hat{H}\rangle$ on a quantum computer, it needs to be expressed using unitary circuits. We have produced the numerical results in this work by decomposing the cost function into Pauli strings. This is not a scalable approach because the number of these strings, in general, scales exponentially with the number of qubits~\cite{motzoi_linear_2017}. When using a Pauli string decomposition for nonlinear problems, it is also necessary to decompose the cost function repeatedly during the optimization process to update the linearized function.
In this section, we show that there is a decomposition into a constant number of circuits using quantum nonlinear processing units (QNPUs)~\cite{lubasch2020variational}, if one can efficiently create the initial state $|\psi_0\rangle$ on a quantum computer. We show that the number of two- or three-qubit gates in these circuits scales linearly with the number of qubits. This is important for the scalability of the algorithm.

The QNPU approach allows us to separate space and time in the cost function, as the propagator $\hat{T}$ is independent of time.
The time part of the $\hat{C}_1$ term~\eqref{eq:suppl-c1} is then given by the sums over $|i\rangle\langle i|$ and $|i+1\rangle\langle i+1|$. In general, the sum over all projectors $|i\rangle\langle i|$ is an identity. However, as these sums run until $N_t-1$, they represent an identity with one missing entry, given by
\begin{subequations}
    \begin{align}
        \sum_{i = 0}^{N_{t}-1} |i\rangle \langle i| &= \hat{I} - |N_t\rangle\langle N_t|,\\
        \sum_{i = 0}^{N_{t}-1} |i+1\rangle \langle i+1| &= \hat{I} - |0\rangle\langle0|.
    \end{align}
\end{subequations}
We can then rewrite the $\hat{C}_1$ term of the cost function as 
\begin{equation}\label{eq:suppl-c1-summed}
    \hat{C}_1 = \hat{I} \otimes \left(\hat{I} - |N_t\rangle\langle N_t|\right) + \hat{T}^\dagger\hat{T} \otimes \left(\hat{I} - |0\rangle\langle0|\right).
\end{equation}

This reduces the complexity of the time part of $\hat{C}_1$ to that of a single projector for both terms. The first term is determined in one circuit by sampling the ansatz and counting the results where the time is not $|N_t\rangle$. The second term can be implemented efficiently by using an ancilla qubit to indicate whether the time was $|0\rangle$, before performing a Hadamard test to measure $\hat{T}^\dagger\hat{T}$, then postselecting the results for times other than $|0\rangle$. In general, the probability of this approaches 1 as $n_t$ is increased.
This circuit is shown in Fig.~\ref{fig:qnpu-circ-c1} in the main text. We propose that one can combine it with a measurement of $\hat{C}_0$ (described below) as shown in Fig.~\ref{fig:suppl-circ-c0-c1}.

The $\hat{C}_2$ term~\eqref{eq:suppl-c2} concerns the shift operator $|i+1\rangle\langle i|$ and its adjoint. This term can be encoded efficiently using an \textit{adder} circuit, such as the one by Lubasch \textit{et al}.~\cite{lubasch2020variational}, with a depth that scales linearly with the number of qubits. For small numbers of qubits one could use the one by Sato \textit{et al}.~\cite{PhysRevA.104.052409} instead.
The shift operator in $\hat{C}_2$ is not periodic, thus it is necessary to filter out the $|N_t\rangle\langle0|$ term. We achieve this by adding a qubit to our time register which we initialize to zero; instead of $|N_t\rangle\langle0|$, there will be an $|N_t\rangle\langle N_t+1|$ term, for which the Hadamard test will measure an overlap of zero. This can be seen in Fig.~\ref{fig:suppl-circ-c2}.

The spatial part of the $\hat{C}_1$ and $\hat{C}_2$ terms is given by the propagator $\hat{T}$, as well as $\hat{T}^\dagger\hat{T}$. For the second-order Taylor approximation used in our main results, $\hat{T}$ contains terms up to the square of the Laplace operator, and $\hat{T}^\dagger\hat{T}$ up to its fourth power.
In general, one can formulate a quantum nonlinear processing unit (QNPU) as given by Lubasch \textit{et al}.~\cite{lubasch2020variational} to calculate any of the terms in the equation. In the propagator $\hat{T}$ to second order, one has the Laplacian $\frac{\partial^2}{\partial x^2}$, the nonlinear term $f\frac{\partial}{\partial x}$, as well as the squares of both and the product of the two with each other. For $\hat{T}^\dagger\hat{T}$, one also needs to implement the third and fourth powers.
The first derivative $\frac{\partial}{\partial x}$ and the Laplacian $\frac{\partial^2}{\partial x^2}$ can be produced by applying an adder circuit in space, implementing the finite difference approximations of either derivative as $(\hat{A}-\hat{I})/\Delta x$ and $(\hat{A} - 2\hat{I} + \hat{A}^\dagger)/(\Delta x)^2$, respectively. For the nonlinearity, one can create a duplicate of the ansatz and do a pointwise multiplication.
The differential operator $\mathcal{L}$ from Eq.~\eqref{eq:diffeq} can then be calculated using
\begin{equation}\label{eq:QNPU}
    \mathcal{L} = \frac{1}{(\Delta x)^2}\left(\hat{A}-2\hat{I}+\hat{A}^\dagger\right) - \beta\mathcal{M}\frac{1}{\Delta x} \left(\hat{A}-\hat{I}\right)\hat{c},
\end{equation}
where $\hat{A}$ represents the adder circuit and $\hat{c}$ the multiplication with a duplicate of the ansatz. One can see this as a diagonal operator with the coefficients of the state on the diagonal. $\mathcal{M}$ is a constant with which the nonlinear term needs to be multiplied, to turn the amplitudes of the quantum state into the function values $f(x,t)$. The value of this can be determined from the norm $\mathcal{N}$ of the initial state divided by the probability to measure $\ket{0}$ time.

\begin{figure*}
\centering
\begin{subfigure}{0.24\textwidth}
    \centering
    {\small (a)}
    \begin{quantikz}[thin lines,row sep=0.1cm,font=\small]
        \lstick{$|0\rangle$} & \gate[2]{U} \gategroup[6,steps=2,style={dashed, rounded corners, red}] \qw & \qw & \qw \\
        \lstick{$|0\rangle$} & \qw & \gate[2]{U} & \qw \\
        \lstick{$|0\rangle$} & \gate[2]{U} & \qw & \qw \\
        \lstick{$|0\rangle$} & \qw & \gate[2]{U} & \qw \\
        \lstick{$|0\rangle$} & \gate[2]{U} & \qw & \qw \\
        \lstick{$|0\rangle$} & \qw & \qw & \qw
    \end{quantikz}
    \phantomcaption\label{figSM:bw-circ}
\end{subfigure}~~
\begin{subfigure}{0.28\textwidth}
    \centering
    {\small (b)}
    \begin{quantikz}[thin lines,row sep=0.1cm,column sep=0.2cm,font=\small]
    \lstick{\ket{0}} & \gate[3]{U} & \qw         & \qw               & \qw         & \qw        \\
    \lstick{\ket{0}} & \qw         & \gate[3]{U} & \qw         & \qw & \qw \\
    \lstick{\ket{0}} & \qw         & \qw         & \gate[3]{U} & \qw & \qw \\
    \lstick{\ket{0}} & \qw         & \qw         & \qw         & \gate[3]{U} & \qw\\ 
    \lstick{\ket{0}} & \qw         & \qw         & \qw         & \qw         &   \qw \\
    \lstick{\ket{0}} & \qw         & \qw         & \qw         & \qw               & \qw
    \end{quantikz}
    \phantomcaption\label{figSM:mps-chi-4}
\end{subfigure}~
\begin{subfigure}{0.45\textwidth}
    \centering
    {\small (c)}
    \begin{quantikz}[thin lines,row sep=0.1cm,font=\small]
    & \gate[2]{U} & \qw         & \gate[2]{U}               & \qw \gategroup[3,steps=2,style={dashed, rounded corners, red}]{}         & \gate[2]{U}
    & \qw & \qw \\
    & \qw         & \gate[2]{U} & \qw & \gate[2]{U} & \qw & \gate[2,style={dashed}]{U} & \qw\\
    & \qw         & \qw         & \qw & \qw & \qw & \qw & \qw
    \end{quantikz}
    \phantomcaption\label{figSM:mps-3q}
\end{subfigure}

\vspace{-1em}
\begin{subfigure}{.38\textwidth}
\centering
{\small (d)}
\begin{quantikz}[thin lines,row sep=0.15cm,column sep=0.2cm,font=\small]
& \ctrl{1} & \gate{R_z} & \gate{R_x} & \gate{R_z} & \qw\\
& \targ{} & \gate{R_z} & \gate{R_x} & \gate{R_z} & \qw\\
\end{quantikz}
\phantomcaption\label{figSM:bw_2qu}
\end{subfigure}~
\begin{subfigure}{.6\textwidth}
\centering
{\small (e)}
\begin{quantikz}[thin lines,row sep=0.15cm,column sep=4pt,inner xsep=-2pt,font=\small]
&\gate[style=dashed]{R_z} & \ctrl{1} & \gate{R_z} & \gate{R_x} & \gate{R_z}
& \ctrl{1} \gategroup[2,steps=2,style={dashed, rounded corners, red}]{$r=3$} & \gate{R_z}
& \ctrl{1} \gategroup[2,steps=4,style={dashed, rounded corners, green}]{$r\geq 2$} & \gate{R_z} & \gate{R_x} & \gate{R_z} & \qw\\
&\qw & \targ{} & \gate{R_z} & \gate{R_x} & \gate{R_z}
& \targ{} & \gate{R_x}
& \targ{} & \gate{R_z} & \gate{R_x} & \gate{R_z} & \qw
\end{quantikz}
\phantomcaption\label{figSM:mps-2q}
\end{subfigure}
\caption{Schematics of a brickwall ansatz with a single layer and a generic MPS ansatz of bond dimension $\chi=4$. (a) One layer of the brickwall ansatz. (b) Generic quantum MPS with $\chi=4$. (c) Three-qubit unitary, with the part taken out for the sparse $\chi=4$ quantum MPS shown in red. The dashed unitary is left out if it overlaps with one of the next block. (d) Two-qubit unitary with one CNOT for the brickwall ansatz. (e) Two-qubit unitary for the MPS ansatz, for a number of CNOTs per block of $r=1...3$. The dashed gate is left out if it overlaps with an $R_z$ gate of the previous block.}
\end{figure*}
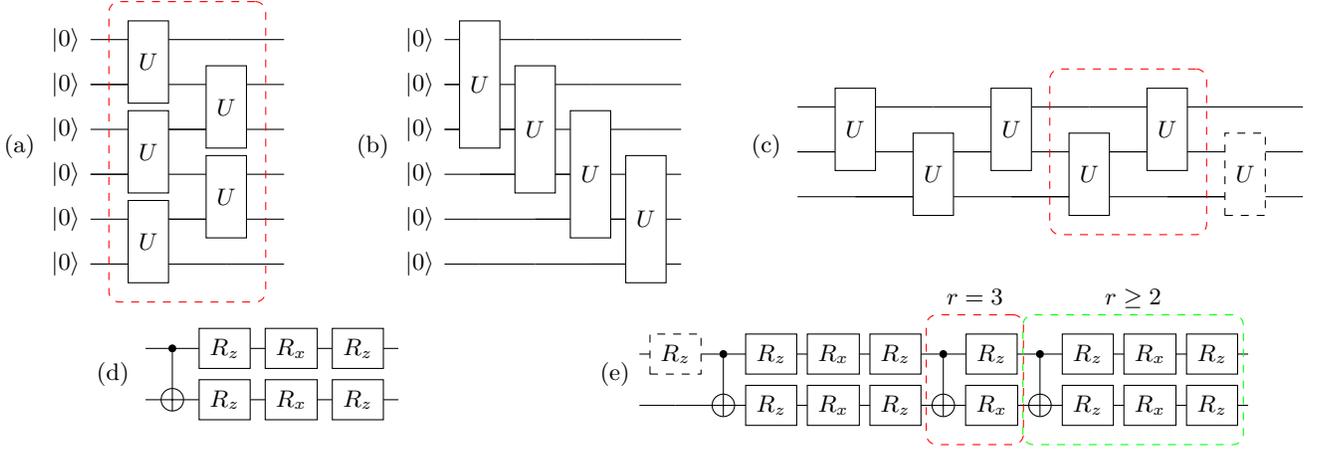

In this QNPU formulation, a complex-valued equation~\eqref{eq:quantum-diff-eq} is being solved, instead of the real-valued Burgers equation, because the pointwise multiplication $\hat{c}$ multiplies with the complex amplitudes of the state. To reproduce the real-valued equation, the state $\ket{\Psi}$ has to be real-valued. This can be ensured using a real-valued ansatz~\cite{sarma_quantum_2024}, or by adding a penalty term to the cost function, such as $\hat{C}_3=c_3[1-\Re(\sum_i\psi_i^2)]$, with $\Re$ the real value, which is zero if all amplitudes are real. A circuit that can be used to measure $\Re(\sum_i\psi_i^2)$ up to a factor $\sqrt{N_tN_x}$ is shown in Fig.~\ref{fig:suppl-qnpu_psi2}.

In the representation of Eq.~\eqref{eq:QNPU}, the propagator $\hat{T}$ can be expressed up to first order in 4 circuits, because one can take $\hat{A}$ and its adjoint together.
The operator $\hat{T}^\dagger\hat{T}$ from the $\hat{C}_1$ can be expressed in 11 circuits up to first order.
The term $\hat{I} \otimes (\hat{I} - |N_t\rangle\langle N_t|)$ from Eq.~\eqref{eq:suppl-c1-summed} can be measured in a single circuit, and the same holds for $\psi^2$ from $\hat{C}_3$ when using a complex-valued ansatz. One additional circuit is necessary to determine the probability to measure the zero time state, which is needed for the normalization constant $\mathcal{M}$. This brings us to a total of 18 circuits for the Burgers equation with a first-order propagator.

The coefficients in front of all of these circuits depend on the Taylor series of the exponential of $\Delta t\,\mathcal{L}$ from Eq.~\eqref{eq:T}, with $\mathcal{L}$ from Eq.~\eqref{eq:QNPU}. For the diffusion term, this means that the coefficients scale with powers of $\Delta t/(\Delta x)^2$, while for the nonlinear term, they scale with $\Delta t/\Delta x$. It is important to make sure that these coefficients do not increase exponentially with the number of space or time qubits. A good scaling can be achieved when $\Delta t\propto\Delta x^2$, i.e., adding two time qubits to the system for every added space qubit. 
Interestingly, this coincides with the stability condition for a PDE that is first order in time and second order in space, which requires that $\Delta t/(\Delta x)^2$ is bounded{~\cite{li_numerical_2017}.

\subsection{Initial state}
The time part of $\hat{C}_0$ is the projector $|0\rangle\langle0|$, and the space part is an identity minus a projector $\hat{I}-|\psi_0\rangle\langle\psi_0|$.
If one can prepare the initial state $|\psi_0\rangle$ as a quantum circuit, it is possible to do a swap test to measure the overlap between the ansatz and this initial state. This results in a linear number of controlled swap gates. One can make this more hardware-efficient using a destructive swap test~\cite{benedetti2021hardware}, which only requires CNOT gates instead of controlled swap gates. However, it is necessary to sample the time register and filter out the time states that are not $|0\rangle$.
Because $\hat{C}_0$ only measures time $|0\rangle$, and the $\hat{T}^\dagger\hat{T}$ term in $\hat{C}_1$ only measures times other than $|0\rangle$, we propose a circuit where one conditionally measures $\hat{C}_0$ or $\hat{C}_1$, based on the state of the time qubits. This is shown in Fig.~\ref{fig:suppl-circ-c0-c1}.

\begin{figure}
    \centering
    \begin{subfigure}{\textwidth}
        \includegraphics[trim={1mm 3mm 0 3mm},clip]{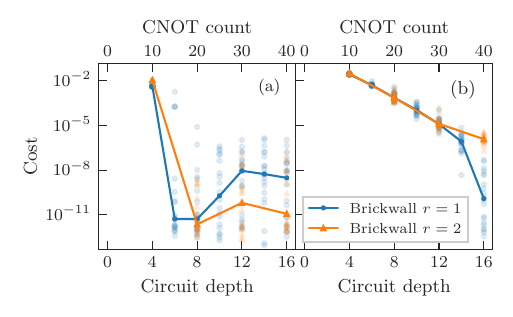}
        \phantomcaption\label{fig:suppl-bw-linear}
        \phantomcaption\label{fig:suppl-bw-nonlinear}
    \end{subfigure}
    \caption{Comparison of the cost $\langle\hat{H}\rangle$ achieved using the brickwall ansatz with $r=1$~or~$2$ CNOTs per block, with a reversed space entanglement structure, and a brickwall ansatz with 2 to 8 layers for $r=1$ and 1 to 4 layers for $r=2$. The lines represent the median of the cost resulting from 20 different random initializations, which are plotted as scattered points. (a) Diffusion. (b)~Burgers equation.}
    \label{fig:suppl-bw}
\end{figure}

This measurement can be simplified if the initial state is a basis vector of the computational basis, or if a basis transformation can be applied such that the initial state can be identified by a single bitstring. In this case, one can sample all qubits and determine the probability of finding time $|0\rangle$ and a spatial state that is not $|\psi_0\rangle$. 
In particular, the $\hat{C}_0$ term for a sine wave as initial condition could be measured using a quantum Fourier transform on the space qubits instead of a swap test.

\section{Additional ansatz circuits}\label{sec:suppl-ansatz}
\begin{figure}
    \centering
    \begin{subfigure}{\textwidth}
        \includegraphics[trim={1mm 3mm 0 3mm},clip]{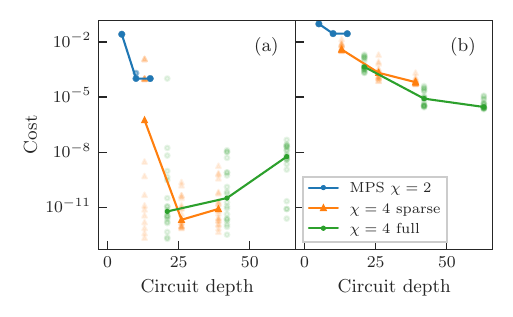}
        \phantomcaption\label{fig:suppl-mps-linear}
        \phantomcaption\label{fig:suppl-mps-nonlinear}
    \end{subfigure}
    \caption{Comparison of the cost $\langle\hat{H}\rangle$ achieved using different MPS ansatz circuits with $\chi$=2 as well as $\chi=4$, with three versions of the two-qubit unitary (three data points on every line) and two versions of the 3-qubit unitary (orange and green lines), with a reversed space entanglement structure. The lines represent the median of the cost resulting from 20 different random initializations, which are plotted as scattered points. (a) Diffusion. (b)~Burgers equation.}
    \label{fig:suppl-mps}
\end{figure}

\subsection{Brickwall ansatz}\label{sec:suppl-bw}

The brickwall ansatz (Fig.~\ref{figSM:bw-circ}) consists of layers of two-qubit unitaries. For the results in the main text, we used a two-qubit unitary consisting of six rotations and one CNOT gate, as given in Fig.~\ref{figSM:bw_2qu}. We have also run the simulations with two CNOTs per block by duplicating this two-qubit unitary, thus creating layers that are twice as deep. The achieved cost for both diffusion and the Burgers equation can be seen in Fig.~\ref{fig:suppl-bw}. We see here, that there is no notable disadvantage in most cases by using less, but deeper, layers.

\begin{figure}
    \centering
    \begin{subfigure}{\textwidth}
        \includegraphics[trim={1mm 3mm 0 3mm},clip]{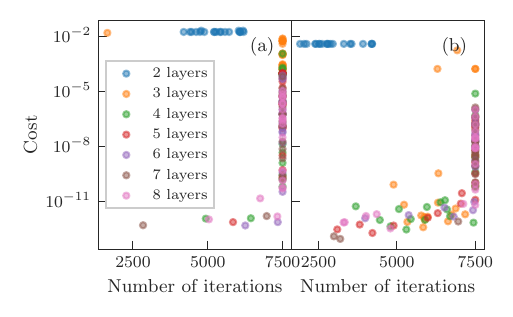}
        \phantomcaption\label{fig:suppl-iter-seq}
        \phantomcaption\label{fig:suppl-iter-seq2}
    \end{subfigure}
    \caption{Cost $\langle\hat{H}\rangle$ versus achieved number of L-BFGS-B iterations for all 20 random initial conditions, for the diffusion problem using a brickwall ansatz with 2 to 8 layers. (a) Sequential entanglement structure. (b) Reversed space.}\label{fig:suppl-iter}
\end{figure}

\begin{figure*}
    \centering
    \begin{subfigure}{.49\textwidth}
        \phantomcaption\label{fig4:burgers_quantinuum}
        \phantomcaption\label{fig4:burgers_ibmq}
        \includegraphics[trim={1mm 3mm 0 3mm},clip]{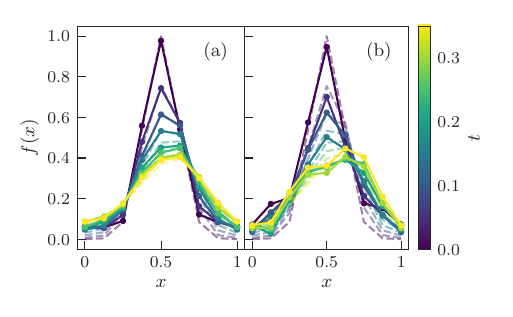}
    \end{subfigure}
    ~
    \begin{subfigure}{.49\textwidth}
        \phantomcaption\label{fig4:diffusion_quantinuum}
        \phantomcaption\label{fig4:diffusion_ibmq}
        \includegraphics[trim={1mm 3mm 0 3mm},clip]{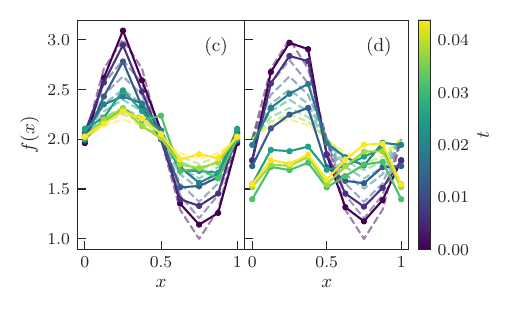}
    \end{subfigure}
    \caption{Solutions to the Burgers and diffusion equation~\eqref{eq:burgers} sampled on NISQ machines. We use $3+3$ qubits ($2^3=8$ points in space and in time) and a reversed space \textit{quantum MPS ansatz} (Sec.~\ref{sec:method_C}) with 26 CNOT gates (Burgers) or 13 CNOT gates (diffusion). (a) Burgers equation, Quantinuum H1-1E emulator. (b) Burgers equation, IBMQ Ehningen. (c) Diffusion equation, Quantinuum H1-1E emulator. (d) Diffusion equation, IBMQ Ehningen.
    The dashed lines correspond to a converged circuit from a noiseless simulator. This noiseless result overlaps with numerical results with an infidelity of $1.89\times10^{-4}$ (Burgers) or $3.2\times10^{-7}$ (diffusion). More details can be found in Sec.~\ref{sec:ibmq-quantinuum}.}
    \label{fig:mps-tomography}
\end{figure*}

\begin{figure*}
    \begin{subfigure}{.49\textwidth}
        \centering
        \includegraphics[trim={1mm 3mm 0 3mm},clip]{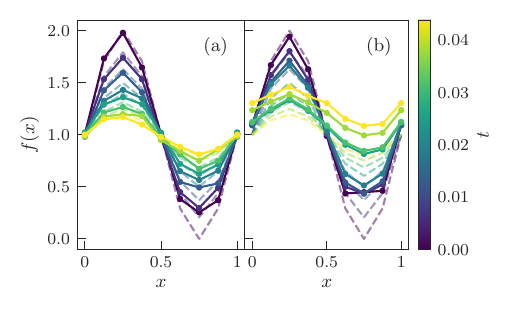}
        \phantomcaption\label{fig:suppl-bw-no-shift-quantinuum}
        \phantomcaption\label{fig:suppl-bw-no-shift-ibmq}
    \end{subfigure}
    ~
    \begin{subfigure}{.49\textwidth}
        \centering
        \includegraphics[trim={1mm 3mm 0 3mm},clip]{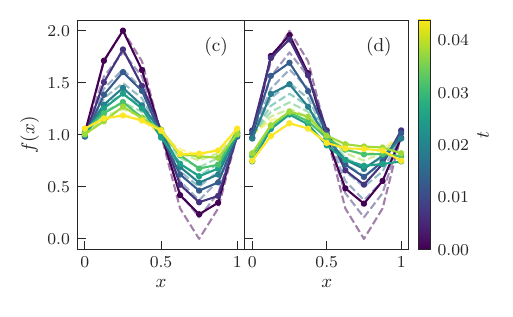}
        \phantomcaption\label{fig:suppl-mps-no-shift-quantinuum}
        \phantomcaption\label{fig:suppl-mps-no-shift-ibmq}
    \end{subfigure}
    \caption{Solutions to the diffusion equation (Eq.~\eqref{eq:burgers} with $\beta=0$), without a shift in the $y$-coordinate, sampled on NISQ machines. We use $3+3$ qubits using (a-b) a brickwall ansatz of 3 layers or (c-d) a quantum MPS ansatz of 13 CNOT gates, and sample the result on (a, c) Quantinuum H1-1E emulator and (b, d) IBMQ Ehningen. The dashed lines correspond to the statevector simulation, which corresponds to results from numerical integration with an infidelity of $1.09\times10^{-6}$ for both cases.}
    \label{fig:suppl-no-shift}
\end{figure*}

\subsection{MPS ansatz}\label{sec:suppl-mps}
We consider an MPS ansatz of bond dimension $\chi=4$ as given in Fig.~\ref{figSM:mps-chi-4}. In general, each of the three-qubit unitaries in this ansatz circuit can be decomposed into six two-qubit unitaries~\cite{barenco1995elementary}. Of these six two-qubit unitaries, the last one overlaps with the first one of the next three-qubit unitary so it is left out, except at the end of the circuit. This means that a structure with five two-qubit unitaries is obtained. To limit the circuit depth, we use a `sparse' MPS with only three two-qubit unitaries, as shown in Fig.~\ref{figSM:mps-3q}. In this circuit, we also vary the complexity of the two-qubit unitary. A generic two-qubit unitary can be composed of 15 rotations and 3 CNOT gates~\cite{shende2004minimal,benedetti2021hardware}. We use three different depths of this circuit as shown in Fig.~\ref{figSM:mps-2q}.
In Fig.~\ref{fig:suppl-mps}, we present the cost as a function of circuit depth
and compare with the `full' $\chi=4$ quantum MPS (with five two-qubit unitaries of varying depths), as well as a $\chi=2$ quantum MPS, to verify that the sparse $\chi=4$ quantum MPS provides a good balance between circuit depth and expressibility.

\section{Optimization protocol}\label{sec:protocols}

For both the diffusion and Burgers equation, we use an optimization procedure where we start with a configuration with a small $D$ and $\beta=0$, and then ramp up either $D$ or $\beta$. This is equivalent to Barison \textit{et al}.~\cite{barison2022variational}, who instead increased the time step in steps to avoid barren plateaus.

We optimize this initial configuration with 2,500 steps of the Adam optimizer~\cite{kingma2014adam}, and all following configurations with at most 2,500 steps L-BFGS-B optimizer~\cite{byrd1995limited}. For the diffusion equation, we start with a diffusion coefficient of $1/8$. After optimizing with Adam, we increase $D$ in steps to $1/4$, $1/2$, and finally $1$, optimizing with L-BFGS-B in-between. For the Burgers equation~\eqref{eq:burgers}, we start with a diffusion coefficient of $D=0.05$ and a nonlinearity coefficient $\beta=0$, so the optimization with Adam can be performed on a linear equation, which is less computationally expensive. We then increase $\beta$ in steps to $1/8$, $1/4$, $1/2$, and finally $1$, optimizing with L-BFGS-B in-between. This means that we end with $\beta=20D$, which is sufficient to observe nonlinear effects.

We chose these optimizers because the Adam optimizer allows us to avoid local minima that can be encountered when starting from a random initial state, whereas the L-BFGS-B optimizer is more local, and thus well suited in cases where the initial state is already reasonably close to the solution. For either optimizer, we use
automatic differentiation~\cite{jones2020efficient}
to determine the gradients of the cost function with respect to all ansatz parameters.

For the L-BFGS-B optimization, we consider the L-BFGS-B optimization converged if the difference in cost between two successive iterations is less than 10 times machine precision. We choose a very low value here to be able to benchmark the cost function, which means that the maximum of 2,500 iterations per value of $D$ is exhausted for most runs of the linear problem.

We run the above protocols for both types of ansatz, with different depths for the brickwall ansatz and different bond dimensions for the quantum MPS. We run each simulation 20 times with random initial conditions. We plot the cost for each random initialization as a function of the number of L-BFGS-B iterations in Fig.~\ref{fig:suppl-iter}. We can see here that the simulations for shallow circuits converge around a cost of $10^{-2}$, while the deeper circuits either exhaust the maximum number of iterations or converge at a cost below $10^{-10}$, which is close to machine precision. Especially for the reversed space structure, there are many runs that converge at a low number of iterations, for all numbers of layers $\geq3$.

For shallow circuits, we find that the cost landscape is irregular, and thus a measurement protocol using different random initializations, then taking the one that leads to the lowest cost, is necessary.
With deep circuits, we find that the optimization will eventually converge if the number of iterations is increased sufficiently, regardless of the initial condition.
As we are focusing on shallow circuits for NISQ devices, we decided to present results from random initializations in this work.

\section{Additional IBMQ and Quantinuum results}
\label{sec:suppl-ibmq}
In the main text, we have shown the final state resulting from noiseless optimization sampled on the IBMQ and Quantinuum machines in Fig.~\ref{fig2:tomography}. In that figure, we used a brickwall ansatz, and the initial condition for the diffusion equation was given by $f(x,t_0)=2+\sin(2\pi x)$.
Here we show similar results with a quantum MPS ansatz in Fig.~\ref{fig:mps-tomography}. We also show results for the diffusion in Fig.~\ref{fig:suppl-no-shift} for both ansatz types
without the upwards shift in the $y$-coordinates, using the initial condition $f(x,t_0)=1+\sin(2\pi x)$. We can see here that both quantum computers have trouble representing amplitudes close to zero.
We see in these figures that the lowest amplitudes are better represented when using a shift, but the overall precision goes down for both quantum computers.

\begin{figure*}
    \centering
    \begin{subfigure}[t]{0.49\textwidth}
        \caption{}
        \includegraphics[trim={0 3mm 0 3mm},clip]{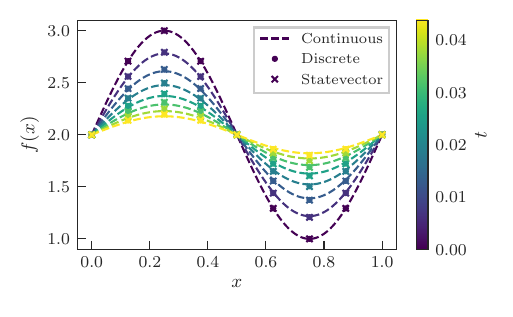}
        \label{fig:diffusion_tomography_continuous}
    \end{subfigure}
    ~
    \begin{subfigure}[t]{0.49\textwidth}
        \caption{}
        \includegraphics[trim={0 3mm 0 3mm},clip]{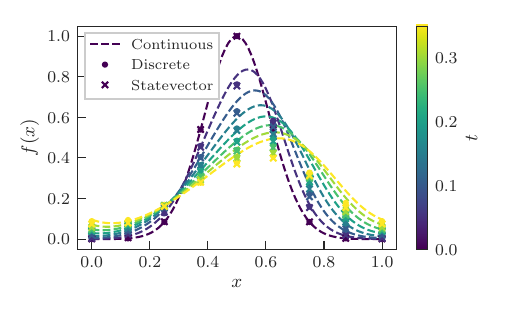}
        \label{fig:burgers_tomography_continuous}
    \end{subfigure}
    \caption{Comparison of the solutions from a noiseless simulator with $3+3$ qubits ($2^3=8$ points in space and in time) to a classical simulation on a fine grid, for (a) diffusion using a brickwall ansatz with 3 layers, (b) Burgers equation using a brickwall ansatz with 4 layers.}
    \label{fig:tomography_continuous}
\end{figure*}
\begin{figure*}
    \centering
    \begin{subfigure}[t]{0.49\textwidth}
        \caption{}
        \includegraphics[trim={0 3mm 0 3mm},clip]{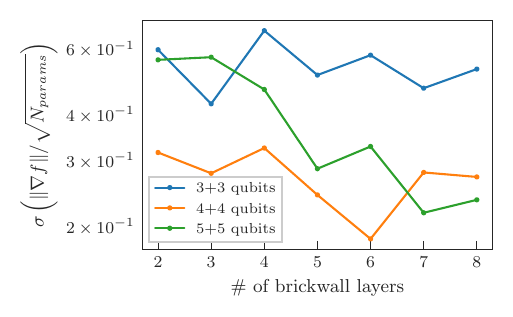}
        \label{fig:diffusion_gradient_stdev_0}
    \end{subfigure}
    ~
    \begin{subfigure}[t]{0.49\textwidth}
        \caption{}
        \includegraphics[trim={0 3mm 0 3mm},clip]{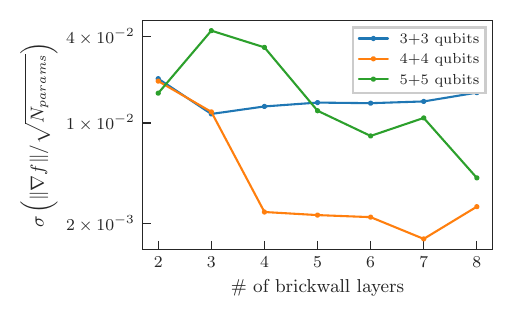}
        \label{fig:diffusion_gradient_stdev_3}
    \end{subfigure}
    \caption{Standard deviation of the gradient norm of the cost function for the diffusion equation with respect to the ansatz parameters, divided by the square root of the number of ansatz parameters, for 20 random initial conditions. (a) We consider the beginning of the first optimization step for $D=0.125$ and random initialization. (b) We consider the beginning of the last optimization step for $D=1$, right after the converged optimization for $D=0.5$.}\label{fig:barren_plateau}
\end{figure*}

It should be noted that the IBMQ results have been produced using readout error mitigation from the M3 library~\cite{nation_scalable_2021}, as well as dynamical decoupling~\cite{souza2012robust} to avoid decoherence during the idle time of the qubits, which is especially relevant for the MPS ansatz. As these methods have not been applied to the results from the Quantinuum platform, there is still room for improvement here. In particular, one could use an ansatz that uses the ZZPhase two-qubit gate defined by $\text{ZZPhase}(\theta)=\exp(-i\frac{\pi\theta}{2}(Z\otimes Z))$, which has an error depending on the phase angle,
instead of the CNOT gate which fully entangles two qubits and is not a native gate of the Quantinuum platform.

In Fig.~\ref{fig:tomography_continuous}, we show the same statevector result that was shown as a dashed line in Fig.~\ref{fig2:tomography}, now combined with a numerical solution on a very fine grid. We see that the error caused by the quantum algorithm is significantly smaller than the discretization error.

\section{Barren plateaus}\label{sec:suppl-barren-plateau}
The presence of barren plateaus is generally indicated by an exponential decrease of the variance of the gradient of the cost function, calculated for different random initializations, as a function of the number of qubits~\cite{mcclean2018barren}.
It has been shown that there is a relation between barren plateaus and ansatz depth~\cite{cerezo2021cost} or ansatz expressibility~\cite{holmes2022connecting} (and a thorough understanding exists for quantum tensor networks~\cite{ZhGa21, CePlLu23} such as the quantum MPS considered in this article), especially for local cost functions. Our cost function is not entirely local, because the shift operators in space and time affect all space or time qubits, which means that one would expect barren plateaus for shallow (or less expressive) ansatz circuits as well. For this reason we calculate the gradient of the cost function during the optimization process for the diffusion equation, using different numbers of qubits and different ansatz depths.

In Fig.~\ref{fig:barren_plateau}, we show the results for two different stages of the optimization: after initializing randomly and after ramping up the diffusion coefficient from $D=0.5$ to $D=1$ (which is the last time that $D$ is increased in the optimization procedure).
Instead of taking the standard deviation of one component of the gradient, we have taken the standard deviation of the gradient norm divided by the square root of the number of parameters to be more rigorous. In this figure, there is no indication of an exponential decrease with the number of qubits.

\section{Noisy VQE}
We perform a preliminary study of the effect of noise on the VQE optimization procedure, by optimizing a $2+2$-qubit system on a noisy simulator with the IBMQ Montreal noise model (another quantum computer with the same 27-qubit chip as IBMQ Ehningen.) We obtained the result from Fig.~\ref{fig:noisy-vqe} after less than 300 iterations. To highlight the effect of noise on the optimization procedure, we have sampled the result state without noise. We see that, despite the noise, optimization is still successful and a state close to the desired solution is found.}

\begin{figure}
    \centering
    \includegraphics{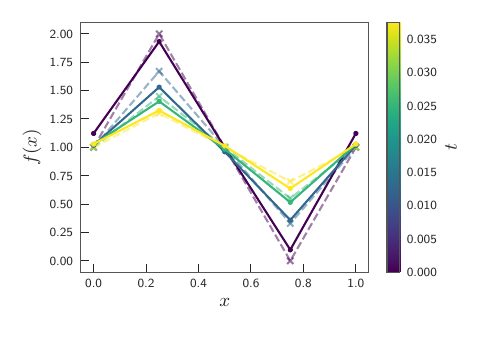}
    \caption{Result of noisy optimization of the diffusion equation on $2+2$ qubits. The optimization was done using the IBMQ Montreal noise model and the result has been plotted without noise. The dashed lines correspond to the solution obtained via numerical integration.}
    \label{fig:noisy-vqe}
\end{figure}

\FloatBarrier
\bibliography{bibliography}

\end{document}